%% file: multiarea_manuscript.tex
\documentclass[english,american,round]{article}

\usepackage[T1]{fontenc}
\usepackage[latin9]{inputenc}
\usepackage[a4paper]{geometry}
\usepackage{fancyhdr}
\usepackage{color}
\usepackage{babel}
\usepackage{array}
\usepackage{prettyref}
\usepackage{float}
\usepackage{mathtools}
\usepackage{url}
\usepackage{amsmath}
\usepackage{graphicx}
\usepackage{setspace}
\usepackage{esint}
\usepackage[authoryear]{natbib}
\usepackage{nomencl}

\providecommand{\makenomenclature}{\makeglossary}
\makenomenclature
\usepackage[unicode=true,
 bookmarks=true,bookmarksnumbered=false,bookmarksopen=false,
 breaklinks=true,pdfborder={0 0 0},backref=false,colorlinks=true]
 {hyperref}
\hypersetup{pdftitle={Full-density multi-scale account of structure and dynamics of macaque visual cortex},
 pdfauthor={Maximilian Schmidt, Rembrandt Bakker, Kelly Shen, Gleb Bezgin, Claus-Christian Hilgetag, Markus Diesmann, and Sacha Jennifer van Albada},
 pdffitwindow, linkcolor=black,citecolor=black,urlcolor=black,filecolor=black}

\makeatletter

\providecommand{\tabularnewline}{\\}

\usepackage{tabularx} 
\usepackage{multirow}
\usepackage{colortbl}
\usepackage{braket}
\usepackage{mathtools}
\usepackage{xr}
\usepackage{rotating}
\usepackage{breakurl}
\usepackage{prettyref}
\usepackage{ifthen}
\newcommand{\arXiv}{true}
\newrefformat{cap}{\hyperref[#1]{Figure \ref{#1}}}
\newrefformat{fig}{\hyperref[#1]{Figure \ref{#1}}}
\newrefformat{sec}{\hyperref[#1]{Sec.~\ref{#1}}}
\newrefformat{sub}{\hyperref[#1]{Sec.~\ref{#1}}}
\newrefformat{tab}{\hyperref[#1]{Table \ref{#1}}}

\newrefformat{supfig}{Figure \ref{#1}}
\newrefformat{suptab}{Table \ref{#1}}

\@ifundefined{definecolor}
 {\usepackage{color}}{}

\definecolor{parametergray}{gray}{0.8}

\usepackage[OT1]{fontenc}

\renewenvironment{abstract}
{\noindent{\normalfont\large\textbf{Summary}}%
\par\vspace{0.5\baselineskip}\noindent}
{\par}

\renewcommand{\@seccntformat}[1]{%
\csname the#1\endcsname\hspace{0.5em}}

\renewcommand{\section}{\@startsection
{section}%
{1}%
{0mm}%
{-\baselineskip}%
{0.5\baselineskip}%
{\normalfont\large\bfseries}}

\renewcommand{\subsection}{\@startsection
{subsection}%
{1}%
{0mm}%
{-\baselineskip}%
{0.5\baselineskip}%
{\normalfont\bfseries}}

\renewcommand{\subsubsection}{\@startsection
{subsubsection}%
{2}%
{1em}%
{-\baselineskip}%
{-\fontdimen2\font plus -\fontdimen3\font minus -\fontdimen4\font}%
{\normalfont\bfseries}}

\usepackage{tocbibind}

\makeatletter
\renewenvironment{table}%
  {\selectfont
  \@float{table}}
  {\end@float}

\ifthenelse{\equal{\arXiv}{true}}{
\renewcommand{\@makecaption}[2]{%
{\small
{\parbox[t]{\linewidth}{%
\renewcommand{\baselinestretch}{1.0}
\vspace{2mm}
\textbf{\textsf{#1}} #2}
}}}
}
{
\usepackage[notabhead,nofighead]{endfloat}

\AtBeginDelayedFloats{%
\sffamily%
\linespread{1}\selectfont%
}

\renewcommand{\@makecaption}[2]{%
{\parbox[t]{\linewidth}{%
\normalsize\renewcommand{\baselinestretch}{1.0}
\vspace{2mm}
\centering
\textbf{\textsf{#1}}
}}}
}

\newcommand{\beginsupplement}{%
        \setcounter{table}{0}
        \renewcommand{\thetable}{S\arabic{table}}%
        \setcounter{figure}{0}
        \renewcommand{\thefigure}{S\arabic{figure}}%
     }

\makeatother

\begin{document}
\include{macros}

\global\long\def\FLN{\mathit{FLN}}
\global\long\def\mm{\:\mathrm{mm}}
\global\long\def\SLN{\mathit{SLN}}

\begin{titlepage}\thispagestyle{empty}\setcounter{page}{0}\pdfbookmark[1]{Title}{TitlePage}

\noindent \begin{center}
\textbf{\huge{}Full-density multi-scale account of structure and dynamics
of macaque visual cortex} \textbf{}
\par\end{center}

\noindent \begin{center}
\textbf{\large{}Maximilian Schmidt$^{1}$, Rembrandt Bakker$^{1,2}$,
Kelly Shen$^{3}$, Gleb Bezgin$^{4}$, Claus-Christian Hilgetag$^{5,6}$,
Markus Diesmann}{\large{}$^{1,7,8}$}\textbf{\large{}, and Sacha Jennifer
van Albada$^{1}$}
\par\end{center}{\large \par}

\vspace{2cm}

\noindent\textrm{$^{1}$}\parbox[t]{15cm}{Institute of Neuroscience
and Medicine (INM-6) and Institute for Advanced Simulation (IAS-6)
and JARA BRAIN Institute I, Jülich Research Centre, Jülich, Germany}\\[3mm]

\noindent\textrm{$^{2}$}\parbox[t]{15cm}{Donders Institute for
Brain, Cognition and Behavior, Radboud University Nijmegen, Netherlands}\\[3mm]

\noindent$^{3}$\parbox[t]{15cm}{Rotman Research Institute, Baycrest,
Toronto, Ontario M6A 2E1, Canada}\\[3mm]

\noindent$^{4}$\parbox[t]{15cm}{McConnell Brain Imaging Centre,
Montreal Neurological Institute, McGill University, Montreal, Canada}\\[3mm]

\noindent$^{5}$\parbox[t]{15cm}{Department of Computational Neuroscience,
University Medical Center Eppendorf, Hamburg, Germany}\\[3mm]

\noindent$^{6}$\parbox[t]{15cm}{Department of Health Sciences,
Boston University, USA}\\[3mm]

\noindent\textrm{$^{7}$}\parbox[t]{15cm}{Department of Psychiatry,
Psychotherapy and Psychosomatics, Medical Faculty, RWTH Aachen University,
Aachen, Germany}\\[3mm]

\noindent$^{8}$\parbox[t]{15cm}{Department of Physics, Faculty
1, RWTH Aachen University, Aachen, Germany}\\[3mm]

\vfill

\noindent$^{*}$Correspondence to:\hspace{1em}\parbox[t]{11cm}{Maximilian
Schmidt\\
Forschungszentrum J\"ulich\\
52425 J\"ulich, Germany\\
\href{mailto:max.schmidt@fz-juelich.de}{max.schmidt@fz-juelich.de}

}

\end{titlepage}
\begin{abstract}
We present a multi-scale spiking network model of all vision-related
areas of macaque cortex that represents each area by a full-scale
microcircuit with area-specific architecture. The layer- and population-resolved
network connectivity integrates axonal tracing data from the CoCoMac
database with recent quantitative tracing data, and is systematically
refined using dynamical constraints. Simulations reveal a stable asynchronous
irregular ground state with heterogeneous activity across areas, layers,
and populations. Elicited by large-scale interactions, the model reproduces
longer intrinsic time scales in higher compared to early visual areas.
Activity propagates down the visual hierarchy, similar to experimental
results associated with visual imagery. Cortico-cortical interaction
patterns agree well with fMRI resting-state functional connectivity.
The model bridges the gap between local and large-scale accounts of
cortex, and clarifies how the detailed connectivity of cortex shapes
its dynamics on multiple scales.

\let\clearpage\relax
\end{abstract}
\ifthenelse{\equal{\arXiv}{true}}{}{

\section{Highlights}

\begin{itemize}
\item A synthesis of anatomical data yields a new population-resolved connectivity
matrix
\item Layer- and area-specific connectivity supports heterogeneous activity
patterns 
\item Spiking activity exhibits increased intrinsic time scales in higher
visual areas
\item The model mimics resting-state fMRI functional connectivity
\end{itemize}

\section{In Brief}

Schmidt et al.\,present a population-resolved full-density spiking
network model of macaque visual cortex integrating a large body of
experimental data. The model combines longer time scales of spiking
activity in higher areas with cortico-cortical interactions mimicking
resting-state activity.

}

\include{introduction}

\include{results}

\include{discussion}

\include{experimental_procedures}

\section{Author Contributions}

Conceptualization: M.D., S.J.v.A., M.S.; Software: M.S., R.B., S.J.v.A.;
Investigation: M.S., S.J.v.A.; Writing - Original Draft: M.S., S.J.v.A.;
Writing - Review \& Editing: M.S., S.J.v.A., R.B., C.-C.H., M.D.;
Resources: R.B., K.S., G.B., C.-C.H.; Funding Acquisition: M.D., M.S.,
S.J.v.A., R.B.; Supervision: S.J.v.A., M.D.

\section{Acknowledgements}

We thank Jannis Schuecker and Moritz Helias for discussions; Helen
Barbas for providing data on neuronal densities and architectural
types; Sarah Beul for discussions on cortical architecture; Kenneth
Knoblauch for sharing his R code for the $SLN$ fit; and Susanne Kunkel
for help with creating \prettyref{fig:construction}D. This work was
supported by the Helmholtz Portfolio Supercomputing and Modeling for
the Human Brain (SMHB), European Union (BrainScaleS, grant 269921
and Human Brain Project, grant 604102), the Jülich Aachen Research
Alliance (JARA), the German Research Council (DFG grants SFB936/A1,Z3
and TRR169/A2), and computing time granted by the JARA-HPC Vergabegremium
and provided on the JARA-HPC Partition part of the supercomputer JUQUEEN
\citep{stephan2015juqueen} at Forschungszentrum Jülich (VSR computation
time grant JINB33).

\renewcommand{\bibname}{References}\bibliographystyle{neuron}

\input{mam_bib.bbl}
\beginsupplement

\include{supplement}

\end{document}

%% file: macros.tex
\selectlanguage{english}%
\global\long\def\taum{\tau_{\text{m}}}

\global\long\def\taur{\tau_{\text{r}}}

\global\long\def\Vr{V_{\mathrm{r}}}

\global\long\def\EL{E_{\mathrm{L}}}

\selectlanguage{american}%
\global\long\def\taus{\tau_{\mathrm{s}}}

\global\long\def\Is{I_{\mathrm{s}}}

\selectlanguage{english}%
\global\long\def\mm{\:\mathrm{mm}}

\selectlanguage{american}%
\global\long\def\mmsquare{\:\mathrm{mm}^{2}}

\selectlanguage{english}%
\global\long\def\mV{\:\mathrm{mV}}

\global\long\def\Hz{\:\mathrm{Hz}}

\global\long\def\ms{\:\mathrm{ms}}

\global\long\def\pA{\:\mathrm{pA}}

\global\long\def\pF{\:\mathrm{pF}}

\global\long\def\cm{C_{\mathrm{m}}}

\global\long\def\cp{C_{\mathrm{p}}}

\global\long\def\ca{C_{\mathrm{a}}}

\global\long\def\mum{\:\mu\mathrm{m}}

\global\long\def\s{\:\mathrm{s}}

\nomenclature{}{}

\global\long\def\taufac{\tau_{\mathrm{fac}}}

\global\long\def\a{\mathrm{a}}

\global\long\def\p{\mathrm{p}}

\global\long\def\e{\mathrm{e}}

\global\long\def\i{\mathrm{i}}

\global\long\def\defeq{\vcentcolon\mathrm{=}}

\selectlanguage{american}%
\global\long\def\nuext{\nu_{\mathrm{ext}}}

\global\long\def\d{\mathrm{d}}

\global\long\def\erf{\mathrm{erf}}

\global\long\def\spikess{\;\mathrm{spikes}/\mathrm{s}}

%% file: introduction.tex
\section{Introduction\label{sec:intro}}

Cortical activity has distinct but interdependent features on local
and global scales, molded by connectivity on each scale. Globally,
resting-state activity has characteristic patterns of correlations
\citep{Vincent07_7140,Fox07,Shen12} and propagation \citep{Mitra14_2374}
between areas. Locally, neurons spike with time scales that tend to
increase from sensory to prefrontal areas \citep{Murray14} in a manner
influenced by both short-range and long-range connectivity \citep{Chaudhuri2015_419}.
We present a full-density multi-scale spiking network model in which
these features arise naturally from its detailed structure.

Models of cortex have hitherto used two basic approaches. The first
models each neuron explicitly in networks ranging from local microcircuits
to small numbers of connected areas \citep{Hill05_1671,Haeusler09_73}.
The second represents the large-scale dynamics of cortex by simplifying
the ensemble dynamics of areas or populations to few differential
equations, such as Wilson-Cowan or Kuramoto oscillators \citep{Deco09_10302,Cabral2011130}.
These models can for instance reproduce resting-state oscillations
at $\sim\!0.1\,\mathrm{Hz}$. \citet{Chaudhuri2015_419} developed
a mean-field multi-area model with a hierarchy of intrinsic time scales
in the population firing rates, relying on a gradient of excitation
across areas.

Cortical processing is not restricted to one or few areas, but results
from complex interactions between many areas involving feedforward
and feedback processes \citep{Lamme98,Rao99}. At the same time, the
high degree of connectivity within areas \citep{Angelucci02_8633,Markov11_1254}
hints at the importance of local processing. Capturing both aspects
requires multi-scale models that combine the detailed features of
local microcircuits with realistic inter-area connectivity. Another
advantage of multi-scale modeling is that it enables testing the equivalence
between population models and models at cellular resolution instead
of assuming it a priori. 

Two main obstacles of multi-scale simulations are now gradually being
overcome. First, such simulations require large resources on high-performance
clusters or supercomputers and simulation technology that uses these
resources efficiently. Recently, important technological progress
has been achieved for the NEST simulator \citep{Kunkel14_78}. Second,
gaps in anatomical knowledge have prevented the consistent definition
of multi-area models. Recent developments in the CoCoMac database
\citep{Bakker12_30} and quantitative axonal tracing \citep{Markov2014_17,Markov14}
have systematized connectivity data for macaque cortex. However, it
remains necessary to use statistical regularities such as relationships
between architectural differentiation and connectivity \citep{Barbas86_415,Barbas97}
to fully specify large cortical network models. Because of these difficulties,
few large-scale spiking network models have been simulated to date,
and existing ones heavily downscale the number of synapses per neuron
\citep{Izhikevich08_3593,Preissl12}, generally affecting network
dynamics \citep{Albada15}. 

We here use realistic numbers of synapses per neuron, building on
a recent model of a $1\mm^{2}$ cortical microcircuit with $\sim\!10^{5}$
neurons \citep{Potjans14_785}. This is the smallest network size
where the majority of inputs per neuron ($\sim\!10,\!000$) is self-consistently
represented at realistic connectivity ($\sim\!10\%$). Nonetheless,
a substantial fraction of synapses originates outside the microcircuit
and is replaced by stochastic input. Our model reduces random input
by including all vision-related areas.

The model combines simple single-neuron dynamics with complex connectivity
and thereby allows us to study the influence of the connectivity itself
on the network dynamics. The connectivity map customizes that of the
microcircuit model to each area based on its architecture and adds
inter-areal connections. By a mean-field method \citep{Schuecker15_arxiv},
we refine the connectivity to fulfill the basic dynamical constraint
of nonzero and non-saturated activity. 

The ground state of cortex features asynchronous irregular spiking
with low pairwise correlations \citep{Ecker10} and low spike rates
$(\sim\!0.1-30\spikess)$ with inhibitory cells spiking faster than
excitatory ones \citep{Swadlow88_1162}. Our model reproduces each
of these phenomena, bridging the gap between local and global brain
models, and relating the complex structure of cortex to its spiking
dynamics.

%% file: results.tex
\section{Results\label{sec:results}}

The model comprises 32 areas of macaque cortex involved in visual
processing in the parcellation of \citet{Felleman91_1}, henceforth
referred to as FV91 (\prettyref{suptab:area_list}). Each area contains
an excitatory and an inhibitory population in each of the layers 2/3,
4, 5 and 6 (L2/3, L4, L5, L6), except area TH, which lacks L4. The
model, summarized in \prettyref{tab:Model-description}, represents
each area by a $1\mm^{2}$ patch.

\subsection{Area-specific laminar compositions}

Neuronal volume densities provided in a different parcellation scheme
are mapped to the FV91 scheme and partly estimated using the average
density of each layer across areas of the same architectural type
(\prettyref{fig:construction}A). Architectural types \citep[Table 4 of ][]{Hilgetag15_submitted}
reflect the distinctiveness of the lamination as well as L4 thickness,
with agranular cortices having the lowest and V1 the highest value.
Neuron density increases with architectural type. When referring to
architectural types, we also use the term `structural hierarchy'.
We call areas like V1 and V2 at the bottom of the structural (or processing)
hierarchy `early', and those near the top `higher' areas.

We find total cortical thicknesses of 14 areas to decrease with logarithmized
overall neuron densities, enabling us to estimate the total thicknesses
of the other $18$ areas (\prettyref{fig:construction}B). Quantitative
data from the literature combined with our own estimates from published
micrographs (\prettyref{suptab:raw_laminar_thicknesses}) determine
laminar thicknesses (\prettyref{fig:construction}C). L4 thickness
relative to total cortical thickness increases with the logarithm
of overall neuron density, which predicts relative L4 thickness for
areas with missing data. Since the relative thicknesses of the other
layers show no notable change with architectural type, we fill in
missing values using the mean of the known data for these quantities
and then normalize the sum of the relative thicknesses to $1$. Layer
thicknesses then follow from relative thickness times total thickness
(see \prettyref{suptab:laminar_thicknesses}).

Finally, for lack of more specific data, the proportions of excitatory
and inhibitory neurons in each layer are taken from cat V1 \citep{Binzegger04}.
Multiplying these with the laminar thicknesses and neuron densities
yields the population sizes (see \nameref{sec:exp_procedures}). 

Each neuron receives synapses of four different origins (\prettyref{fig:construction}D).
In the following, we describe how the counts for these synapse types
are computed (details in \nameref{sec:exp_procedures}).

\subsection{Scalable scheme of local connectivity}

We assume constant synaptic volume density across areas \citep{Harrison02}.
Experimental values for the average indegree in monkey visual cortex
vary between $2,\!300$ \citep{OKusky82_278} and $5,\!600$ \citep{Cragg1967}
synapses per neuron. We take the average ($3,\!950$) as representative
for V1, resulting in a synaptic density of $8.3\cdot10^{8}\:\frac{\textrm{synapses}}{\textrm{mm}^{3}}$.

The microcircuit model of \citet{Potjans14_785} serves as a prototype
for all areas. The indegrees are a defining characteristic of this
local circuit, as they govern the mean synaptic currents. We thus
preserve their relative values when customizing the microcircuit to
area-specific neuron densities and laminar thicknesses. The connectivity
between populations is spatially uniform. The connection probability
averages an underlying Gaussian connection profile over a disk with
the surface area of the simulated area, separating simulated local
synapses (type I) formed within the disk from non-simulated local
synapses (type II) from outside the disk (\prettyref{fig:construction}D,
E). In retrograde tracing experiments, \citet{Markov11_1254} found
the fraction of labeled neurons intrinsic to each injected area ($FLN_{\mathrm{i}}$)
to be approximately constant, with a mean of $0.79$. We translate
this to numbers of synapses by assuming that the proportion of synapses
of type I is $0.79$ for realistic area size. For the $1\mmsquare$
model areas, we obtain an average proportion of type I synapses of
$0.504$.

\subsection{Layer-specific heterogeneous cortico-cortical connectivity}

\begin{figure}
\begin{centering}
\includegraphics[scale=0.8]{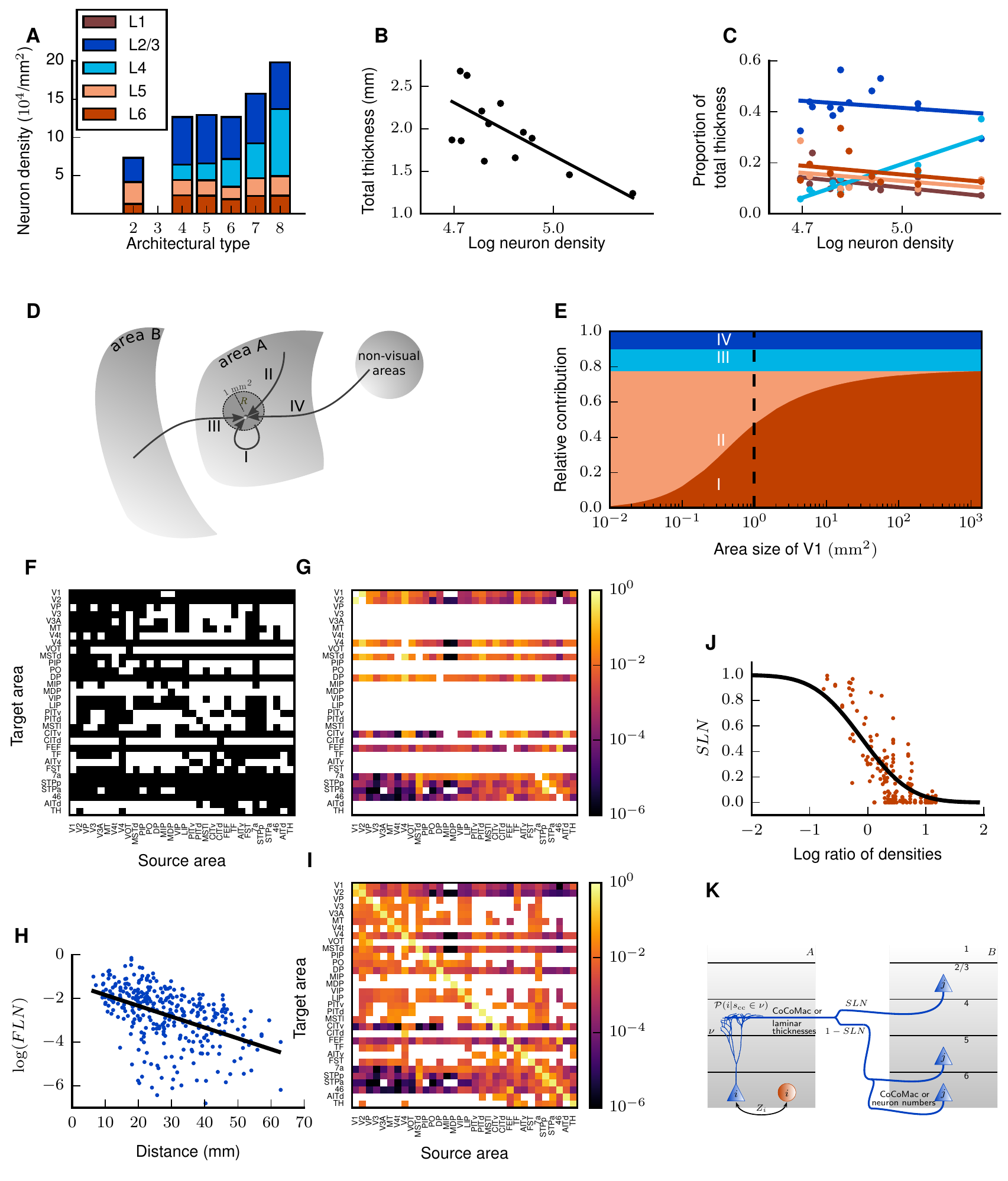}
\par\end{centering}

\caption{\textbf{Construction principles of the model.} (A) Laminar neuron
densities for the architectural types in the model. Type 2, here corresponding
only to area TH, lacks L4. In the model, L1 contains synapses but
no neurons. Data provided by H. Barbas and C. Hilgetag (personal communication)
and linearly scaled up to account for undersampling of cells by NeuN
staining relative to Nissl staining as determined by repeat measurements
of 11 areas. (B) Total thickness vs. logarithmized overall neuron
density and linear least-squares fit \textrm{($r=-0.7,\:p=0.005$).
}(C) Relative laminar thickness (see \prettyref{suptab:raw_laminar_thicknesses})
vs. logarithmized overall neuron density and linear least-squares
fits (L1: $r=-0.51,\,p=0.08$, L2/3: $r=-0.20,\,p=0.52$, L4: $r=0.89,\:p=0.0001$;
L5: $r=-0.31,\:p=0.36$, L6: $r=-0.26,\,p=0.43$). Total cortical
thicknesses $D(A)$ and overall neuron densities for $14$ areas provided
by H. Barbas and C. Hilgetag (personal communication), measured by
Nissl staining for the 11 areas mentioned above and for 3 areas by
NeuN staining and linearly scaled up to account for undersampling.
The data partially overlap with \citet{Hilgetag15_submitted}. (D)
Scheme of the different types of connections to each neuron. I: Simulated
intra-area synapses, II: Intra-area synapses from outside the $1\protect\mmsquare$
patch modeled as Poisson sources, III: Simulated cortico-cortical
synapses, IV: Synapses from subcortical and non-visual cortical areas
modeled as Poisson sources. (E) Relative contributions to indegrees
in V1 for increasing cortical surface area covered by the model. Type
I synapses increase at the cost of random input at type II synapses.
Numbers of type III and IV synapses stay constant. The dashed line
indicates the $1\protect\mmsquare$ surface area used here. (F) Binary
connectivity from CoCoMac. Black, existing connections; white, absent
connections. (G)\textbf{ }Fractions of labeled neurons ($FLN$) from
\citet{Markov2014_17} mapped from their parcellation scheme (M132)
to that of \citet{Felleman91_1}. (H)\textbf{ }Connection densities
decay exponentially with inter-area distance. Black line, linear regression
with $\log\left(FLN\right)=\log\left(C\right)-\lambda d$ ($C=0.045,\lambda=0.11,p=10^{-19}$;
cf.\,\prettyref{eq:EDR}). (I)\textbf{ }Area-level connectivity of
the model, based on data in panels F-H, expressed as relative indegrees
for each target area. (J) Fraction of source neurons in supragranular
layers ($SLN$) vs. logarithmized ratio of the overall neuron densities
of the two areas. $SLN$ from \citet{Markov14}, neuron densities
from \citet{Hilgetag15_submitted}. Black curve, fit using a beta-binomial
generalized linear model \prettyref{eq:SLN_GLM} ($a_{0}=-0.152\text{, }a_{1}=-1.534\text{, }\phi=0.214$).
(K)\textbf{ }Illustration of the procedure for distributing synapses
across layers. Source neuron $j$ from area $B$ sends an axon to
layer $v$ of area $A$ where a cortico-cortical synapse $s_{\mathrm{CC}}$
is formed at the dendrite of neuron $i$. The procedure is detailed
in \nameref{sec:exp_procedures}. See \prettyref{eq:cc_syn_distribution}
for the formal definitions. }
\label{fig:construction}
\end{figure}

We treat all cortico-cortical connections as originating and terminating
in the $1\mmsquare$ patches, ignoring their spatial divergence and
convergence. Two areas are connected if the connection is in CoCoMac
(\prettyref{fig:construction}F) or reported by \citet{Markov2014_17}.
For the latter we assume that the average number of synapses per labeled
neuron is constant across projecting areas (\prettyref{fig:construction}G).
To estimate missing values, we exploit the exponential decay of connectivity
with distance \citep{ErcseyRavasz13}. We first map the data from
its native parcellation scheme (M132) to the FV91 scheme (see \nameref{sec:exp_procedures})
and then perform a least-squares fit (\prettyref{fig:construction}H).
Combining the binary information on the existence of connections with
the connection densities gives the area-level connectivity matrix
(\prettyref{fig:construction}I).

Next, we distribute synapses between the populations of each pair
of areas (\prettyref{fig:construction}K). The pattern of source layers
is based on CoCoMac, if laminar data is available. Fractions of supragranular
labeled neurons ($SLN$) from retrograde tracing experiments yield
proportions of projecting neurons in supra- and infragranular layers
\citep{Markov14}. To predict missing values, we exploit a sigmoidal
relation between the logarithmized ratios of cell densities of the
participating areas and the $SLN$ of their connection (as suggested
by \citealt{Beul15_arxiv}; \prettyref{fig:construction}J). Following
\citet{Markov14}, we use a generalized linear model for the fit and
assume a beta-binomial distribution of source neurons. Since \citet{Markov14}
do not distinguish infragranular layers further into L5 and L6, we
use the more detailed laminar patterns from CoCoMac for this purpose,
if available. We exclude L4 from the source patterns, in line with
anatomical observations \citep{Felleman91_1}, and approximate cortico-cortical
connections as purely excitatory \citep{Salin95_107,Tomioka07_526}.

We base termination patterns on anterograde tracing studies collected
in CoCoMac, if available, or on a relationship between source and
target patterns (see \nameref{sec:exp_procedures}). Since neurons
can receive synapses in different layers on their dendritic branches,
we use laminar profiles of reconstructed cell morphologies \citep{Binzegger04}
to relate synapse to cell-body locations. Despite the use of a point
neuron model, we thus take into account the layer specificity of synapses
on the single-cell level. In contrast to laminar synapse distributions,
the resulting laminar distributions of target cell bodies are not
highly distinct between feedforward and feedback projections.

\subsection{Brain embedding}

Inputs from outside the scope of our model, i.e., white-matter inputs
from non-cortical or non-visual cortical areas and gray-matter inputs
from outside the $1\,\mathrm{mm^{2}}$ patch, are represented by Poisson
spike trains. Corresponding numbers of synapses are not available
for all areas, and laminar patterns of external inputs differ between
target areas \citep{Felleman91_1,Markov14}. Therefore, we determine
the total number of external synapses onto an area as the total number
of synapses minus those of type I and III, and distribute them with
equal indegree for all populations.

\subsection{Refinement of connectivity by dynamical constraints}

\begin{figure}
\includegraphics{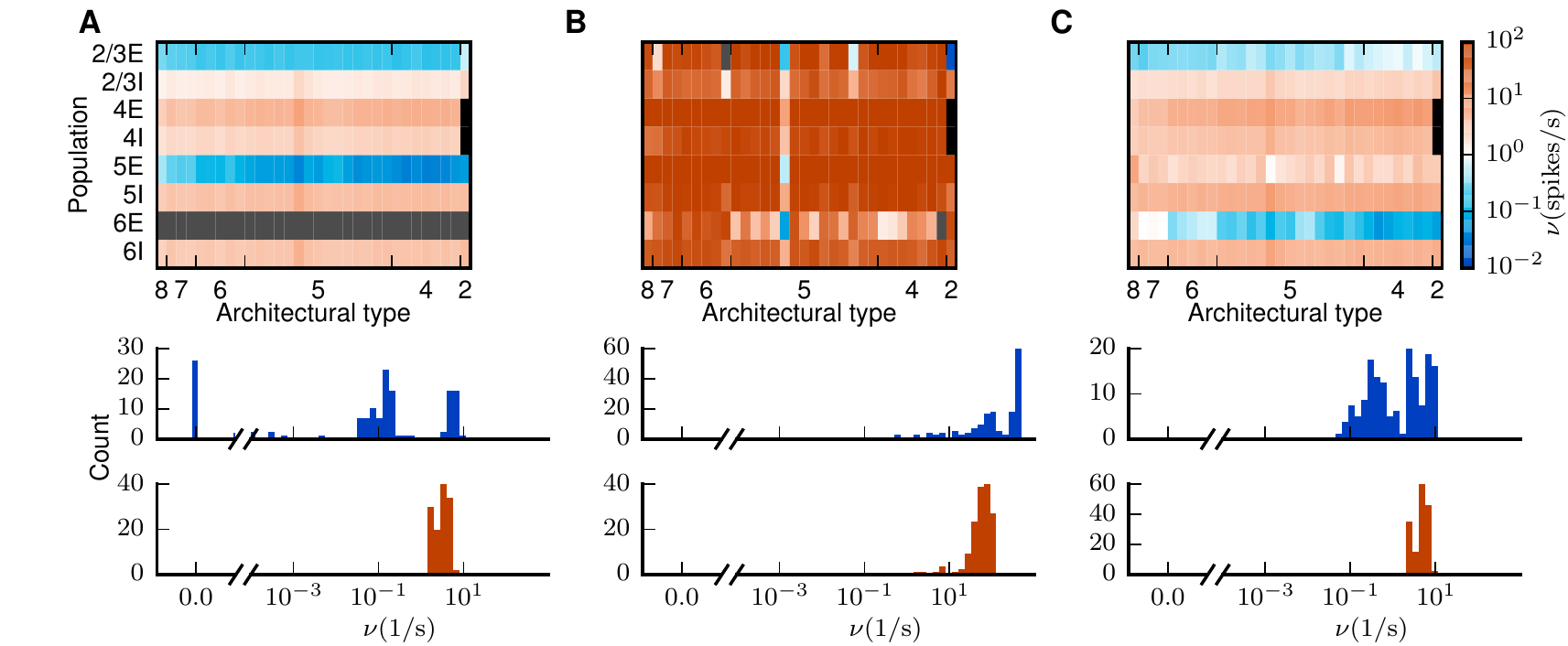}\caption{\textbf{Attractors of the network. }Top row: Firing rates of simulations
with $g=-16$, $\nu_{\mathrm{ext}}=10\:\mathrm{spikes/s}$, $\kappa=1$
(A), $g=-16$, $\nu_{\mathrm{ext}}=10\:\mathrm{spikes/s}$, $\kappa=1.125$
(B), and $g=-11$, $\nu_{\mathrm{ext}}=10\:\mathrm{spikes/s}$, $\kappa=1.125$
with the modified connectivity matrix (C). The color bar holds for
all three panels. Areas are ordered according to their architectural
type along the horizontal axis from V1 (type 8) to TH (type 2) and
populations are stacked vertically. The two missing populations 4E
and 4I of area TH are marked in black and firing rates $<10^{-2}\protect\Hz$
in gray. Bottom row: Histogram of population-averaged firing rates
for excitatory (red) and inhibitory (blue) populations. The horizontal
axis is split into linear- (left) and log-scaled (right) ranges.}
\label{fig:bistability}
\end{figure}

Parameter scans based on mean-field theory \citep{Schuecker15_arxiv}
and simulations reveal a bistable activity landscape with two coexisting
stable fixed points. The first has reasonable firing rates except
for populations 5E and 6E, which are nearly silent (\prettyref{fig:bistability}A),
while the second has excessive rates (\prettyref{fig:bistability}B)
in almost all populations. Depending on the parameter configuration,
either the low-activity fixed point has a sufficiently large basin
of attraction for the simulated activity to remain near it, or fluctuations
drive the network to the high-activity fixed point. To counter this
shortcoming, we define an additional parameter $\kappa$ which increases
the external drive onto 5E by a factor $\kappa=K_{\mathrm{ext,5E}}/K_{\mathrm{ext}}$
compared to the external drive of the other cell types. Since the
rates in population 6E are even lower, we increase the external drive
to 6E by a slightly larger factor than that to 5E. When applied directly
to the model, even a small increase in $\kappa$ already drives the
network into the undesired high-activity state (\prettyref{fig:bistability}B).
Using the stabilization procedure described in \citet{Schuecker15_arxiv},
we derive targeted modifications of the connectivity within the margins
of uncertainty of the anatomical data, with an average relative change
in total indegrees (summed over source populations) of $11.3\%$ (\prettyref{supfig:connectivity-pops}B).
This allows us to increase $\kappa$ while retaining the global stability
of the low-activity state. In the following, we choose $\kappa=1.125$,
which gives $K_{\mathrm{6E,\mathrm{ext}}}/K_{\mathrm{ext}}=1.417$
and the external inputs listed in \prettyref{suptab:external-input},
and $g=-11,\:\nu_{\mathrm{ext}}=10\spikess$, yielding reasonable
firing rates in populations 5E and 6E (\prettyref{fig:bistability}C).
In total, the $4.13$ million neurons of the model are interconnected
via $2.42\cdot10^{10}$ synapses.

The stabilization renders the intrinsic connectivity of the areas
more heterogeneous. Cortico-cortical connection densities similarly
undergo small changes, but with a notable reduction in the mutual
connectivity between areas 46 and FEF. For more details on the connectivity
changes, see \citet{Schuecker15_arxiv}.

\subsection{Community structure of anatomy relates to functional organization}

\begin{figure}
\includegraphics{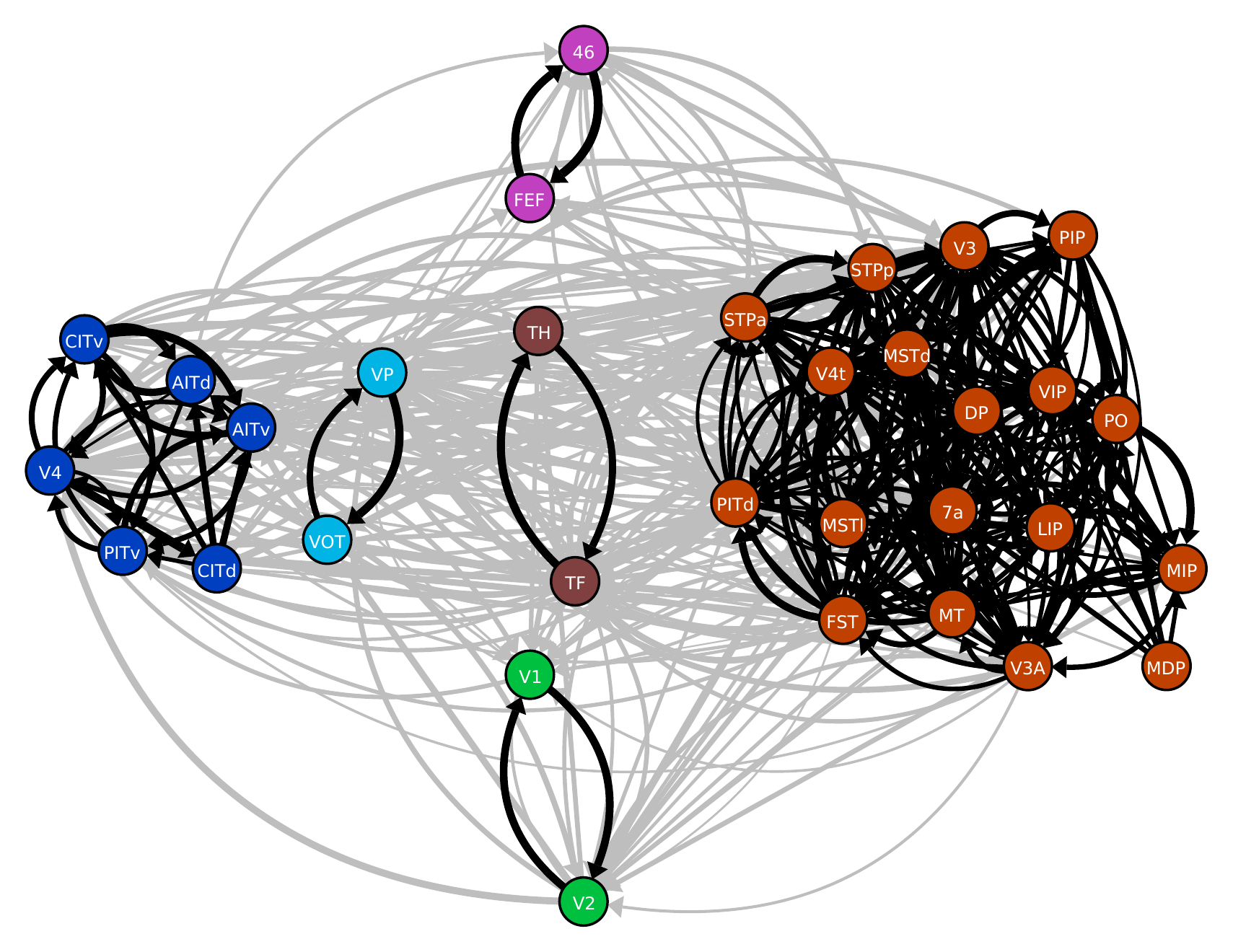}\caption{\textbf{Community structure of the model.} Clusters in the connectivity
graph, indicated by the color of the nodes: Early visual areas (green),
dorsal stream areas (red), areas VP and VOT (light blue), ventral
stream (dark blue), parahippocampal areas (brown), and frontal areas
(purple). Black, connections within clusters; gray, connections between
clusters. Line thickness encodes logarithmized outdegrees. Only edges
with relative outdegree$>10^{-3}$ are shown.}
\label{fig:map_equation}
\end{figure}

We test if the stabilized network retains known organizing principles
by analyzing the community structure in the weighted and directed
graph of area-level connectivity. The map equation method \citep{Rosvall10}
reveals 6 clusters (\prettyref{fig:map_equation}). We test the significance
of the corresponding modularity $Q=0.32$ by comparing with $1000$
surrogate networks conserving the total outdegree of each area by
shuffling its targets. This yields $Q=-0.02\pm0.03$, indicating the
significance of our clustering. The community structure reflects anatomical
and functional properties of the areas. Two large clusters comprise
ventral and dorsal stream areas, respectively. Ventral area VOT is
grouped with early visual area VP. Early sensory areas V1 and V2 form
a separate cluster, as well as parahippocampal areas TH and TF. The
two frontal areas FEF and 46 form the last cluster. Nonetheless, the
clusters are heavily interconnected (\prettyref{fig:map_equation}).
The basic separation into ventral and dorsal clusters matches that
found in the connection matrix of \citet{Felleman91_1} \citep{Hilgetag00}
containing about half of the connections present in our weighted connectivity
matrix, but there are also important differences. For instance, our
clustering groups areas STPa, STPp, and 7a with the dorsal instead
of the ventral stream, better matching the scheme described by \citet{Nassi09_360}.

\subsection{Area- and population-specific activity in the resting state}

\begin{figure}
\includegraphics{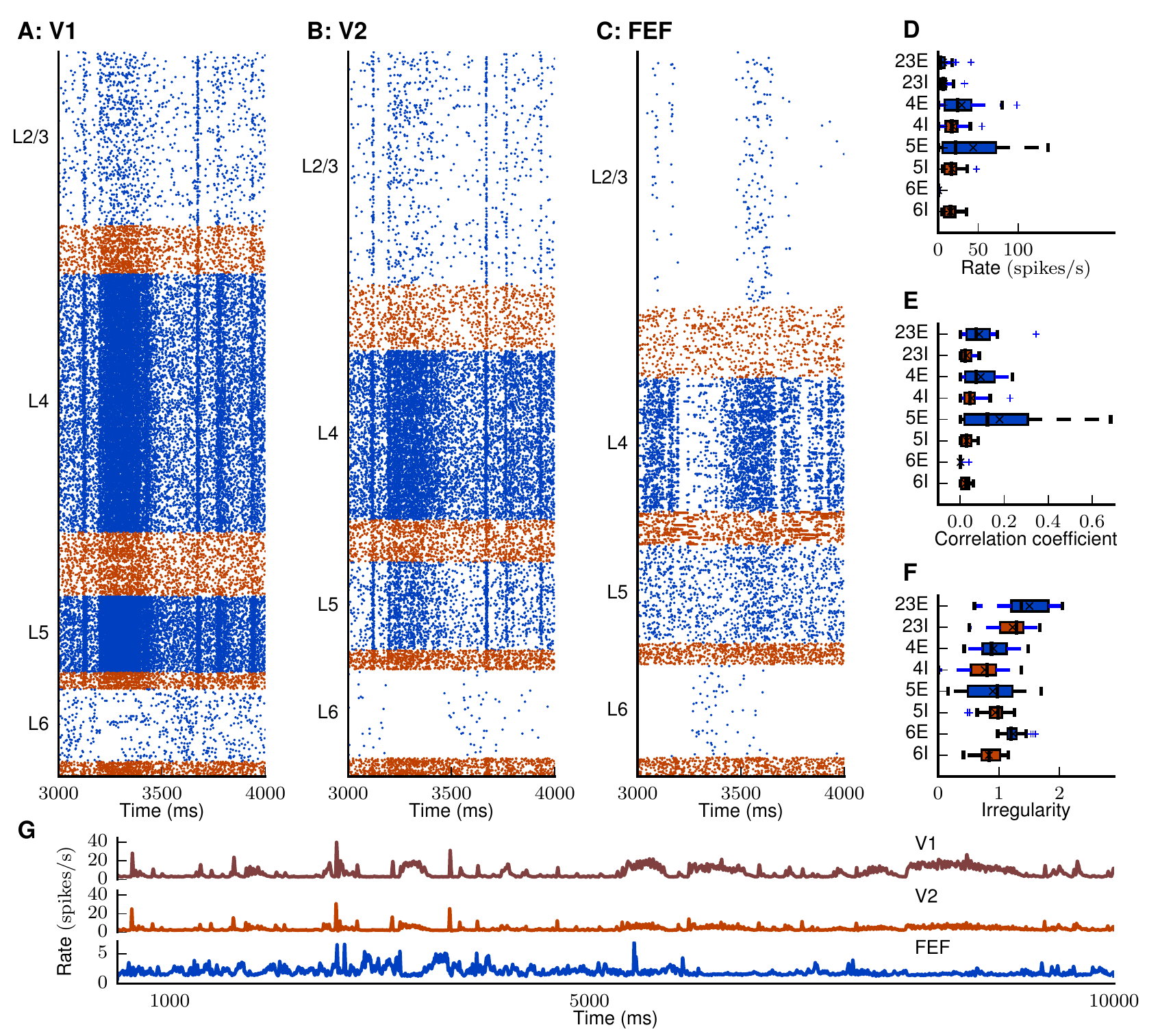}

\caption{\textbf{Resting state of the model.} (A-C) Raster plot of spiking
activity of $3\,\%$ of the neurons in area V1 (A), V2 (B), and FEF
(C). (D-F) Spiking statistics across areas and populations shown as
area-averaged box plots. (D) Population-averaged firing rates. (E)
Average pairwise correlation coefficients of spiking activity. (F)
Irregularity measured by revised local variation $LvR$ \citep{Shinomoto09_e433}
averaged across neurons. (G) Area-averaged firing rates.}
\label{fig:ground_state}
\end{figure}

The model with cortico-cortical synaptic weights equal to local weights
displays a reasonable ground state of activity but no substantial
inter-area interactions (\prettyref{supfig:ground_state_lambda1}).
To control these interactions, we scale cortico-cortical synaptic
weights $w_{\mathrm{cc}}$ onto excitatory neurons by a factor $\lambda=J_{\mathrm{cc}}^{\mathcal{E}}/J$
and provide balance by increasing the weights $J_{\mathrm{cc}}^{\mathrm{\mathcal{I}}}$
onto inhibitory neurons by twice this factor, $J_{\mathrm{cc}}^{\mathcal{I}}=\lambda_{\mathcal{I}}\lambda J=2\lambda J$.
In the following, we choose $\lambda=1.9$. Simulations yield irregular
activity with plausible firing rates (\prettyref{fig:ground_state}A-C).
Irregularly occurring population bursts of different lengths up to
several seconds arise from the asynchronous baseline activity (\prettyref{fig:ground_state}G)
and propagate across the network. The firing rates differ across areas
and layers and are generally low in L2/3 and L6 and higher in L4 and
L5, partly due to the cortico-cortical interactions (\prettyref{fig:ground_state}D).
The overall average rate is $14.6\spikess$. Inhibitory populations
are generally more active than excitatory ones across layers and areas
despite the identical intrinsic properties of the two cell types.
However, the strong participation of L5E neurons in the cortico-cortical
interaction bursts causes these to fire more rapidly than L5I neurons.
 Pairwise correlations are low throughout the network (\prettyref{fig:ground_state}E).
Excitatory neurons are more synchronized than inhibitory cells in
the same layer, except for L6. Spiking irregularity is close to that
of a Poisson process across areas and populations, with excitatory
neurons consistently firing more irregularly than inhibitory cells
(\prettyref{fig:ground_state}F). Higher areas exhibit bursty spiking,
as illustrated by the raster plot for area FEF (\prettyref{fig:ground_state}C).

\subsection{Intrinsic time scales increase with structural hierarchy}

\begin{figure}
\includegraphics{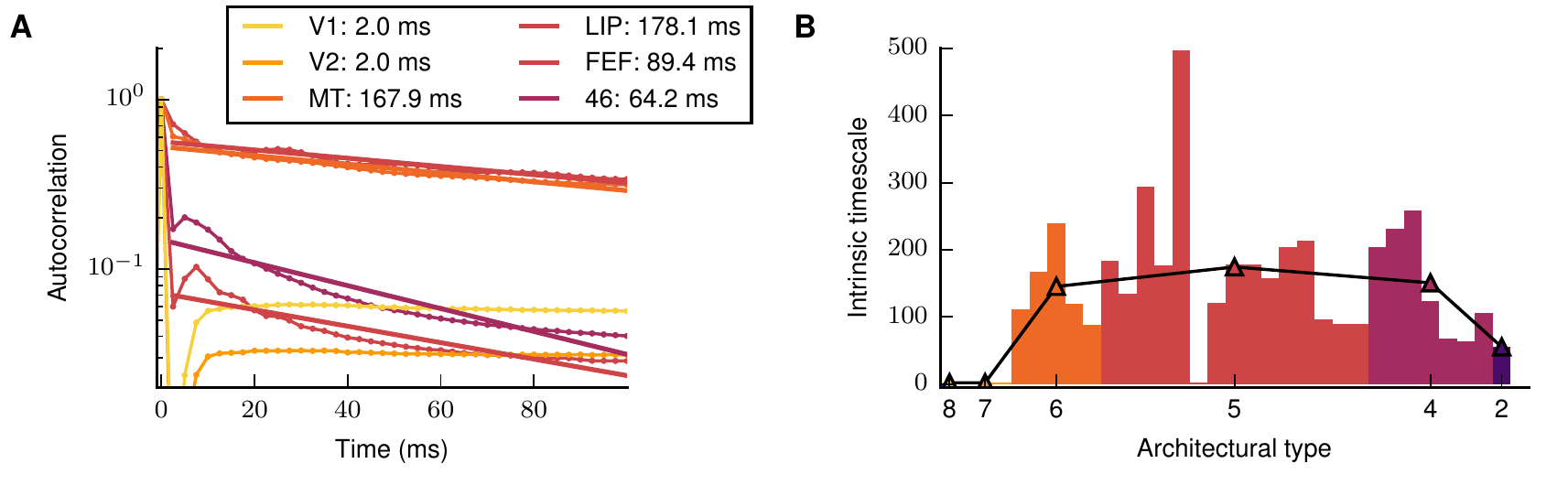}\caption{\textbf{Intrinsic time scales}. (A) Dots, autocorrelation function
averaged across all populations in each area (colors in legend). Lines,
exponential fit $f(t)=A\cdot\exp[-t/\tau]$. (B) Intrinsic time scales
of areas vs. architectural type (color-coded). Average time scale
per architectural type indicated by triangles and overall trend by
black curve. Area MDP (architectural type 5) has a time scale of $2\protect\ms$
because it is uncoupled from other areas due to the lack of incoming
connections.}
\label{fig:intrinsic_timescales}
\end{figure}

We tested whether the model accounts for the hierarchical trend in
intrinsic time scales observed in macaque cortex \citep{Murray14}.
Indeed, autocorrelation width in the model increases from early visual
to higher areas. In early visual areas including V1, the autocorrelation
decays with $\tau<2.5\ms$, indicating near-Poissonian spiking (\prettyref{fig:intrinsic_timescales}A).
In higher areas, autocorrelations are broader with decay times $\sim\!10^{2}\ms$.
The long time scales reflect bursty spike patterns of single-neuron
activity (\prettyref{fig:ground_state}), caused by the low neuron
density in higher areas and thus high indegrees due to the constant
synaptic density. A simulation with equal intrinsic and long-range
synaptic weights that showed no significant interactions yielded near-Poissonian
spiking in all areas (\prettyref{supfig:ground_state_lambda1}), showing
that the cortico-cortical interactions elicit the increased time scales.
Area 46, which overlaps with lateral prefrontal cortex studied by
\citet{Murray14}, shows a shorter time scale compared to the experimental
data. However, in line with \citet{Murray14}, we find the time scale
of area LIP to exceed that of MT, albeit by a small amount. 

\subsection{Structural and hierarchical directionality of spontaneous activity}

\begin{figure}
\includegraphics{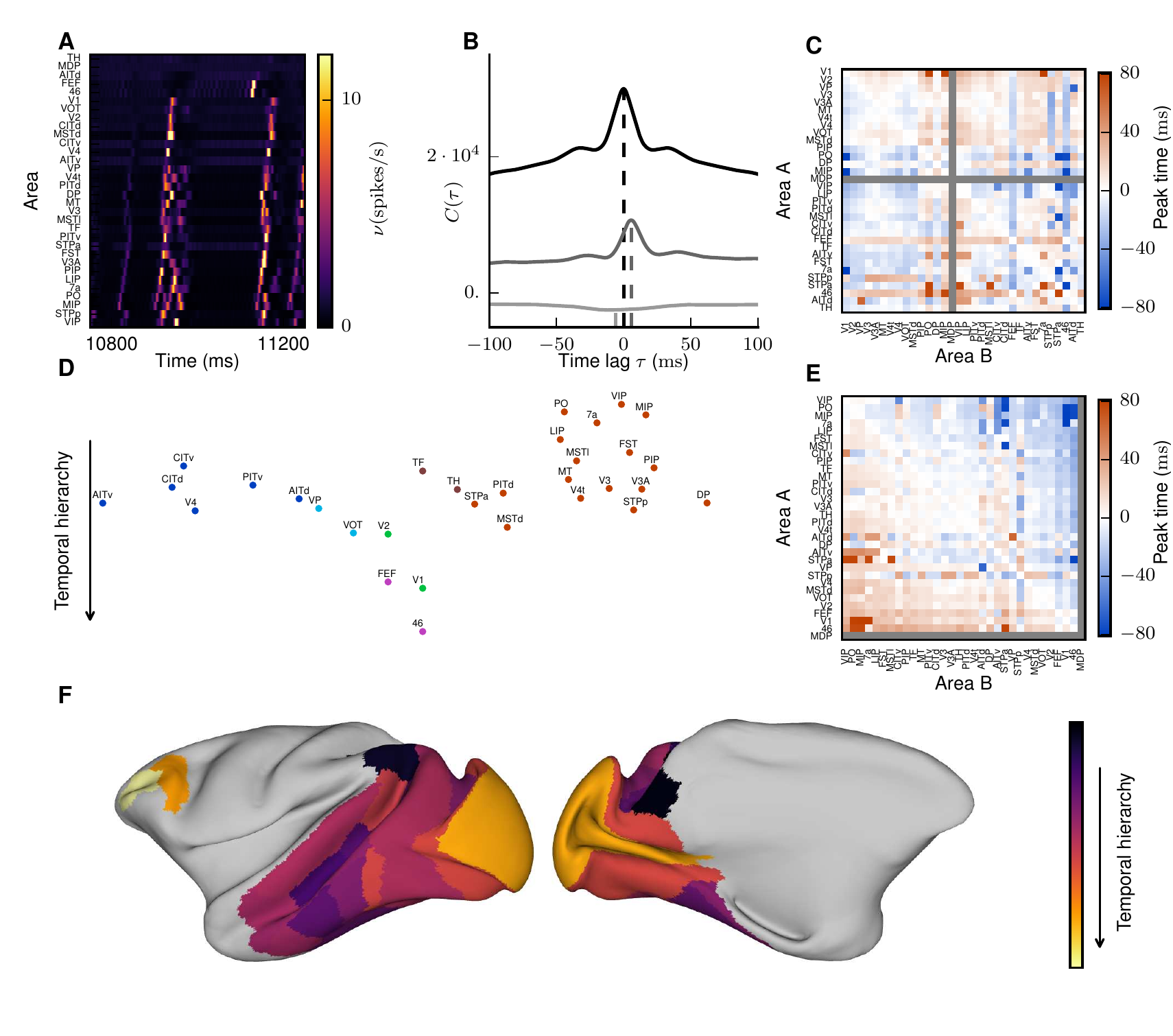}\caption{\textbf{Temporal hierarchy. }(A)\textbf{ }Area-averaged firing rates
for a sample period, with areas ordered according to the onset of
increased activity from $t>1250\protect\ms$. (B) Covariance functions
of the area-averaged firing rates of V1 with areas V2 (gray) and FEF
(light gray), and auto-covariance function of V1 (black). Dashed lines
mark selected time lags, detected by a wavelet smoothing algorithm
(see \nameref{sec:exp_procedures}). (C) Matrix of time lags of the
correlation function for all pairs of areas. Area MDP was neglected
because it has only outgoing connections to but no incoming connections
from other visual areas according to CoCoMac. (D) Temporal hierarchy.
Colors correspond to the map equation clustering (cf. \prettyref{fig:map_equation}).
Areas are horizontally arranged to avoid visual overlap. (E) Peak
position matrix with areas in hierarchical order. (F) Lateral (left)
and medial (right) view on the left hemisphere of an inflated macaque
cortical surface showing the order in which areas are preferentially
activated. Created with the ``view/map 3d surface'' tool on \protect\url{http://scalablebrainatlas.incf.org}.}
\textbf{\label{fig:correlation_hierarchy}}
\end{figure}

To investigate inter-area propagation, we determine the temporal order
of spiking (\prettyref{fig:correlation_hierarchy}A) based on the
correlation between areas. We detect the location of the extremum
of the correlation function for each pair of areas (\prettyref{fig:correlation_hierarchy}B)
and collect the corresponding time lags in a matrix (\prettyref{fig:correlation_hierarchy}C).
In analogy to structural hierarchies based on pairwise connection
patterns \citep{Reid09_611}, we look for a temporal hierarchy that
best reflects the order of activations for all pairs of areas (see
\nameref{sec:exp_procedures}). The result (\prettyref{fig:correlation_hierarchy}D)
places parietal and temporal areas at the beginning and early visual
as well as frontal areas at the end. The first and second halves of
the time series yield qualitatively identical results (\prettyref{supfig:validation_hierarchy}).
\prettyref{fig:correlation_hierarchy}E shows the consistency of the
hierarchy with the pairwise lags. To quantify the goodness of the
hierarchy, we counted the pairs of areas for which it indicates a
wrong ordering. The number of such violations is $190$ out of $496$,
well below the $230\pm12\,\text{(SD)}$ violations obtained for $100$
surrogate matrices, created by shuffling the entries of the original
matrix while preserving its antisymmetric character. This indicates
that the simulated temporal hierarchy reflects nonrandom patterns.
The propagation is mostly in the feedback direction not only in terms
of the structural hierarchy, but also spatially: activity starts in
parietal regions, and spreads to the temporal and occipital lobes
(\prettyref{fig:correlation_hierarchy}F). However, activity troughs
in frontal areas follow peaks in occipital activity and thus appear
last.

\subsection{Emerging interactions mimic experimental functional connectivity}

\begin{figure}
\includegraphics{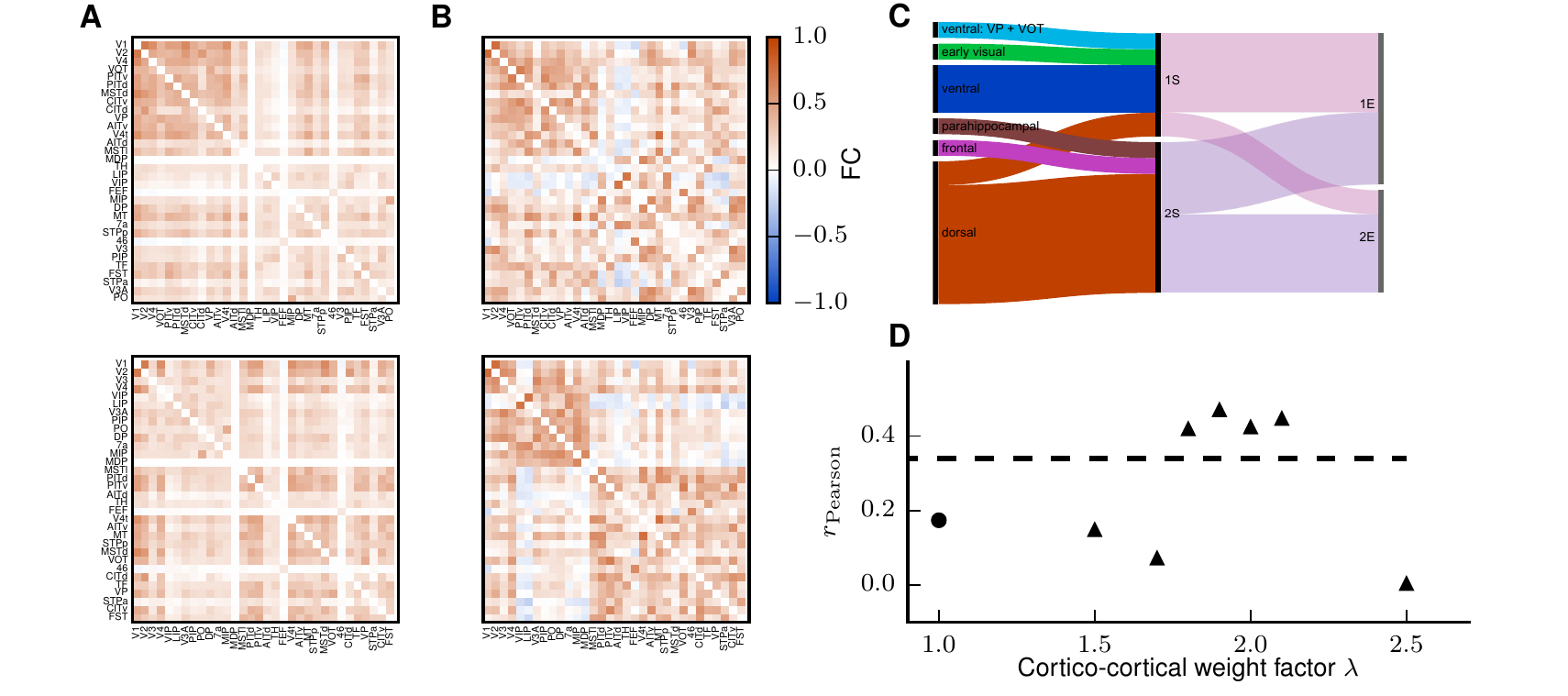}\caption{\textbf{Inter-area interactions. }(A) Simulated functional connectivity
(FC) for $\lambda=1.9$ measured by the zero-time lag correlation
coefficient of synaptic input currents. (B) FC from macaque resting-state
fMRI (see \nameref{sec:exp_procedures}). Areas are ordered according
to a clustering with the Louvain algorithm \citep{Blondel2008} applied
to the simulated data (top row) and to the experimental data (bottom
row), respectively (see \nameref{sec:exp_procedures}).\textbf{ }(C)\textbf{
}Alluvial diagram showing the differences in the clusters for the
structural connectivity (left), the simulated FC (center) and the
experimentally measured FC (right). (D) Pearson correlation coefficient
of simulated FC vs. experimentally measured FC for varying $\lambda$
with $\lambda_{\mathcal{I}}=2$ (triangles) and $\lambda_{\mathcal{I}}=1$
(dot, cf. \prettyref{supfig:ground_state_lambda1}). Dashed line,
Pearson correlation coefficient of structural connectivity vs. experimentally
measured FC.}
\label{fig:interactions}
\end{figure}

We compute the area-level functional connectivity (FC) based on the
synaptic input current to each area, which has been shown to be more
comparable to the BOLD fMRI than the spiking output \citep{Logothetis01}.
The FC matrix exhibits a rich structure, similar to experimental resting-state
fMRI (\prettyref{fig:interactions}A, B, see \nameref{sec:exp_procedures}
for details). In the simulation, frontal areas 46 and FEF are more
weakly coupled with the rest of the network, but the anticorrelation
with V1 is well captured by the model (\prettyref{supfig:anticorrelation}).
Moreover, area MDP sends connections to, but does not receive connections
from other areas according to CoCoMac, limiting its functional coupling
to the network. Louvain clustering \citep{Blondel2008}, an algorithm
optimizing the modularity of the weighted, undirected FC graph \citep{Newman04},
yields two modules for both the simulated and the experimental data.
The modules from the simulation differ from those of the structural
connectivity and reflect the temporal hierarchy shown in \prettyref{fig:correlation_hierarchy}C.
Cluster 1S merges early visual with ventral and two dorsal regions
with average level in the temporal hierarchy of $\overline{h}=0.47\pm0.13\,\text{(SD)}$.
Cluster 2S contains mostly temporally earlier areas ($\overline{h}=0.33\pm0.25\,\text{(SD)}$)
merging parahippocampal with dorsal but also frontal areas. The experimental
module 2E comprises only dorsal areas, while 1E consists of all other
areas including also eight dorsal areas.

The structural connectivity of our model shows higher correlation
with the experimental FC ($r_{\mathrm{Pearson}}=0.34$) than the binary
connectivity matrices from both a previous \citep{Shen15_5579} and
the most recent release of CoCoMac ($r_{\mathrm{Pearson}}=0.20)$,
further validating our weighted connectivity matrix. For increasing
weight factor $\lambda$, the correlation between simulation and experiment
improves (\prettyref{fig:interactions}D). For $\lambda=1$, areas
interact weakly, resulting in low correlation between simulation and
experiment (\prettyref{supfig:ground_state_lambda1}). For intermediate
cortico-cortical connection strengths, the correlation of simulation
vs. experiment exceeds that between the structural connectivity and
experimental FC (\prettyref{fig:interactions}C), indicating the enhanced
explanatory power of the dynamical model. From $\lambda=2$ on, the
network is prone to switch to the high-activity state (\prettyref{supfig:LA-HA-switching}).
Thus, the highest correlation ($r_{\mathrm{Pearson}}=0.47$ for $\lambda=1.9$)
occurs just below the onset of a state in which the model visits both
the low-activity and high-activity attractors.

%% file: discussion.tex
\section{Discussion}

In this work, we present a full-density spiking multi-scale network
model of all vision-related areas of macaque cortex. An updated connectivity
map at the level of areas, layers, and neural populations defines
its structure. Simulations of the network on a supercomputer reveal
good agreement with multi-scale dynamical properties of cortex and
supply testable hypotheses. Consistent with experimental results,
the local structure of areas supports higher firing rates in inhibitory
than in excitatory populations, and a laminar pattern with low firing
rates in layers 2/3 and 6 and higher rates in layers 4 and 5. When
cortico-cortical interactions are substantial, the network shows dynamic
characteristics reflecting both local and global structure. Individual
cells spike irregularly with increasing intrinsic time scales along
the visual hierarchy and activity propagates in the feedback direction.
Functional connectivity in the model agrees well with that from resting-state
fMRI and yields better predictions than the structural connectivity
alone. These features are direct consequences of the multi-scale structure
of the network.

The structure of the model integrates a wide range of anatomical data,
complemented with statistical predictions. The cortico-cortical connectivity
is based on axonal tracing data collected in a new release of CoCoMac
\citep{Bakker12_30} and recent quantitative and layer-specific retrograde
tracing \citep{Markov14,Markov2014_17}. We fill in missing data using
relationships between laminar source and target patterns \citep{Felleman91_1,Markov14},
and statistical dependencies of cortico-cortical connectivity on
distance \citep{ErcseyRavasz13} and architectural differentiation
\citep{Beul15_arxiv,Hilgetag15_submitted}, an approach for which
\citet{Barbas86_415,Barbas97} laid the groundwork. The use of axonal
tracing results avoids the pitfalls of diffusion MRI data, which strongly
depend on tractography parameters and are unreliable for long-range
connections \citep{Thomas14}. Direct comparison of tracing and tractography
data moreover reveals that tractography is particularly unreliable
at fine spatial scales, and tends to underestimate cortical connectivity
\citep{Calabrese15_bhv121}. 

Our model customizes the microcircuit of \citet{Potjans14_785} based
on the specific architecture of each area, taking into account neuronal
densities and laminar thicknesses. A stabilization procedure \citep{Schuecker15_arxiv}
further diversifies the internal circuitry of areas. Neuronal densities
in the model decrease up the structural hierarchy, in line with an
observed caudal-to-rostral gradient \citep{Charvet2015_147}. Combined
with a constant synaptic volume density \citep{OKusky82_278,Cragg1967}
this yields higher indegrees up the hierarchy. This trend matches
an increase in dendritic spines per pyramidal neuron \citep{Elston00_RC117,Elston00_RC95,Elston11},
also used in a recent multi-area population rate model \citep{Chaudhuri2015_419}.
The local connectivity can be further refined using additional area-specific
data.

We find total cortical thickness to decrease with logarithmized total
neuron density. Similarly, total thicknesses from MR measurements
decrease with architectural type \citep{Wagstyl15_241}, which is
known to correlate strongly with cell density \citep{Hilgetag15_submitted}.
In our data set, total and layer 4 thickness are also negatively correlated
with architectural type, but these trends are less significant than
those with logarithmized neuron density. Laminar and total cortical
thicknesses are determined from micrographs, which has the drawback
that this covers only a small fraction of the surface of each cortical
area. For absolute but not relative thicknesses, another caveat is
potential shrinkage and obliqueness of sections. It has also been
found that relative laminar thicknesses depend on the sulcal or gyral
location of areas, which is not offset by a change in neuron densities
\citep{Hilgetag06_146}. However, regressing our relative thickness
data against cortical depth of the areas registered to F99 revealed
no significant trends of this type (\prettyref{supfig:thickness_vs_depth}).
Laminar thickness data are surprisingly incomplete, considering that
this is a basic anatomical feature of cortex. In future, more systematic
estimates from anatomical studies or MRI may become available. Total
thicknesses have already recently been measured across cortex \citep{Calabrese15_408,Wagstyl15_241},
and could complement the dataset used here covering 14 of the 32 areas.
However, when computing numbers of neurons, using histological data
may be preferable, because shrinkage effects on neuronal densities
and laminar thicknesses partially cancel out.

In the model, we statistically assign synapses to target neurons
based on anatomical reconstructions \citep{Binzegger04}. On the
target side, this yields similar laminar cell-body distributions for
feedforward and feedback projections despite distinct laminar synapse
distributions, mirroring findings in early visual cortex of mouse
\citep{DePasquale11}. Prominent experimental results on directional
differences in communication patterns are based on LFP, ECoG and MEG
recordings \citep{Kerkoerle14_14332,Bastos15_390,Michalareas2016_384},
which mostly reflect synaptic inputs. In future, these findings may
be integrated into the stabilization procedure to better capture such
differential interactions. While this is expected to enhance the distinction
between average connection patterns for feedforward and feedback projections,
known anatomical patterns suggest that a substantial fraction of individual
pairs of areas deviate from a simple rule \citep{Felleman91_1,Krumnack10,Bakker12_30}.
The cortico-cortical connectivity may be further refined by incorporating
the dual counterstream organization of feedforward and feedback connections
\citep{Markov14}, or by taking into account different numbers of
inter-area synapses per neuron in feedforward and feedback directions
\citep{Rockland04_387}.

In the resulting connectivity, we find multiple clusters reflecting
the anatomical and functional partition of visual cortex into early
visual areas, ventral and dorsal streams, parahippocampal and frontal
areas, showing that the model construction yields a meaningful network
structure. Moreover, the graded structural connectivity of the model
agrees better with the experimentally measured resting-state activity
than the binary connectivity from CoCoMac.

The network exhibits an asynchronous, irregular ground state across
the network with population bursts due to inter-area interactions.
Population firing rates differ across layers and inhibitory rates
are generally higher than excitatory ones, in line with experimental
findings \citep{Swadlow88_1162,Fujisawa08_823,Sakata09_404}. This
can be attributed to the connectivity, because excitatory and inhibitory
neurons are equally parametrized and excitatory neurons receive equal
or stronger external stimulation compared to inhibitory ones. Laminar
activity patterns vary across areas due to their customized structure
and cortico-cortical connectivity.

Intrinsic single-cell time scales in the model are short in early
visual areas and long in higher areas, on the same order of magnitude
as found experimentally \citep{Murray14}. The long time scales in
higher areas are related to bursty firing associated with the high
indegrees in these areas, but only occur in the presence of cortico-cortical
interactions. Thus, the model predicts that the pattern of intrinsic
time scales has a multi-scale origin. Systematic differences in synaptic
composition across cortical regions and layers \citep{Zilles04,Hawrylycz2012}
may also contribute to the experimentally observed pattern of time
scales.

Inter-area interactions in the model are mainly mediated by population
bursts of different lengths. The degree of synchrony accompanying
inter-area interactions in the brain is as yet unclear. Obtaining
substantial cortico-cortical interactions with low synchrony may be
possible with finely structured connectivity and reduced noise input.
When neurons are to a large extent driven by a noisy external input,
a smaller percentage of their activity is determined by intrinsic
inputs, which can decrease their effective coupling \citep{Aertsen90}.
One way of reducing the external drive while preserving the mean network
activity may be for the drive to be attuned to the intrinsic connectivity
\citep{Marre09_14596}. Stronger intrinsic coupling while maintaining
stability may be achieved for instance by introducing specific network
structures such as synfire chains \citep{Diesmann99} or other feedforward
structures, subnetworks, or small-world connectivity \citep{Jahnke14_030701};
population-specific patterns of short-term plasticity \citep{Sussillo07_4079};
or fine-tuned inhibition between neuronal groups \citep{Hennequin14_1394}.

The synchronous population events propagate stably across multiple
areas, predominantly in the feedback direction. The systematic activation
of parietal before occipital areas in the model is reminiscent of
EEG findings on information flow during visual imagery \citep{Dentico2014_237}
and the top-down propagation of slow waves during sleep \citep{Massimini04_6862,Nir11_153,Sheroziya14_8875}.
Our method for determining the order of activations is similar to
one recently applied to fMRI recordings \citep{Mitra14_2374}. It
could be extended to distinguish between excitatory and inhibitory
interactions like those we observe between V1 and frontal areas (\prettyref{supfig:anticorrelation}).
In the network, cortico-cortical projections target both excitatory
and inhibitory populations, with the majority of synapses terminating
on excitatory cells. Stronger cortico-cortical synapses to enhance
inter-area interactions require increased balancing of cortico-cortical
inputs to preserve network stability. This is similar to the ``handshake''
mechanism in the microcircuit model of \citet{Potjans14_785} where
interlaminar projections provide network stability by their inhibitory
net effect.

The pattern of simulated interactions between areas resembles fMRI
resting-state activity. The agreement between simulation and experiment
peaks at intermediate coupling strength, where synchronized clusters
also emerged most clearly in earlier models \citep{Zhou06,Deco12_3366}.
Furthermore, optimal agreement occurs just below a transition to a
state where the network switches between attractors, supporting evidence
that the brain operates in a slightly subcritical regime \citep{Deco12_3366,Priesemann14_80}.

Time series of spiking activity reveal broad-band transmission between
areas on time scales up to several seconds. The low-frequency part
of these interactions is comparable to fMRI data, which describes
coherent fluctuations on the order of seconds. The long time scales
in the model activity may be caused by eigenmodes of the effective
connectivity that are close to instability \citep{Bos2015_arxiv}
or non-orthogonal \citep{Hennequin_12}. A potential future avenue
for research would be to distinguish between such network effects
and other sources of long time scales such as NMDA and $\mathrm{GABA_{B}}$
transmission, neuromodulation, or adaptation effects. 

For tractability, the model represents each area as a $1\mmsquare$
patch of cortex. True area sizes vary from $\sim\!3$ million cells
in TH to $\sim\!300$ million cells in V1 for a total of around $8\cdot10^{8}$
neurons in one hemisphere of macaque visual cortex, a model size that
with recent advances in simulation technology \citep{Kunkel14_78}
already fits on the most powerful supercomputers available today.
Approaching this size would reduce the negative effects of downscaling
\citep{Albada15}. 

Overall, our model elucidates multi-scale relationships between cortical
structure and dynamics, and can serve as a platform for the integration
of new experimental data, the creation of hypotheses, and the development
of functional models of cortex.

%% file: experimental_procedures.tex
\section{Experimental procedures\label{sec:exp_procedures}}

\begin{table}[H]
\begin{tabular}{@{\hspace*{1mm}}p{3cm}@{}|@{\hspace*{1mm}}p{12.2cm}}
\hline 
\multicolumn{2}{>{\color{white}\columncolor{black}}c}{\textbf{A: Model summary}}\tabularnewline
\hline 
\textbf{Populations} &  254 populations: 32 areas (\prettyref{suptab:area_list}) with eight
populations each (area TH: six) \tabularnewline
\textbf{Topology} & ---\tabularnewline
\textbf{Connectivity} & area- and population-specific but otherwise random\tabularnewline
\textbf{Neuron model } & leaky integrate-and-fire (LIF), fixed absolute refractory period
(voltage clamp)\tabularnewline
\textbf{Synapse model} & exponential postsynaptic currents\tabularnewline
\textbf{Plasticity} & ---\tabularnewline
\textbf{Input} & independent homogeneous Poisson spike trains \tabularnewline
\textbf{Measurements} & spiking activity\tabularnewline
\end{tabular}

\begin{tabular}{@{\hspace*{1mm}}p{3.cm}@{}@{\hspace*{1mm}}p{2.8cm}@{}@{\hspace*{1mm}}p{4.8cm}@{}@{\hspace*{1mm}}p{4.4cm}}
\hline 
\multicolumn{4}{>{\color{white}\columncolor{black}}c}{\textbf{B: Populations}}\tabularnewline
\textbf{Type} & \textbf{Elements} & \textbf{Number of populations} & \textbf{\hspace{5mm}Population size}\tabularnewline
Cortex & LIF neurons & 32 areas with eight populations each (area TH: six), two per layer & \hspace{5mm}$N$ (area- and population-specific)\tabularnewline
\end{tabular}

\begin{tabular}{@{\hspace*{1mm}}p{3cm}@{}|@{\hspace*{1mm}}p{12.2cm}}
\multicolumn{2}{>{\color{white}\columncolor{black}}c}{\textbf{C: Connectivity}}\tabularnewline
\textbf{Type} & source and target neurons drawn randomly with replacement (allowing
autapses and multapses) with area- and population-specific connection
probabilities\tabularnewline
\textbf{Weights} & fixed, drawn from normal distribution with mean $J$ and standard
deviation \mbox{$\delta J = 0.1 J$}; 4E to 2/3E increased by factor
$2$ \citep[cf. ][]{Potjans14_785}; weights of inhibitory connections
increased by factor $g$; excitatory weights $<0$ and inhibitory
weights $>0$ are redrawn; cortico-cortical weights onto excitatory
and inhibitory populations increased by factor $\lambda$ and $\lambda_{\mathcal{I}}\lambda$,
respectively\tabularnewline
\textbf{Delays} & fixed, drawn from Gaussian distribution with mean $d$ and standard
deviation \mbox{$\delta d = 0.5 d$}; delays of inhibitory connections
factor $2$ smaller; delays rounded to the nearest multiple of the
simulation step size $h=0.1\,\mathrm{ms}$, inter-areal delays drawn
from a Gaussian distribution with mean $d=s/v_{\mathrm{t}}$, with
distance $s$ and transmission speed $v_{\mathrm{t}}=3.5\,\mathrm{m/s}$
\citep{Girard01}; and standard deviation $\delta d=d/2$, distances
determined as described in \nameref{sec:Supplement-Exp}, delays $<0.1\,\mathrm{ms}$
before rounding are redrawn\tabularnewline
\end{tabular}

\begin{tabular}{@{\hspace*{1mm}}p{3cm}@{}|@{\hspace*{1mm}}p{12.2cm}}
\multicolumn{2}{>{\color{white}\columncolor{black}}c}{\textbf{D: Neuron and synapse model}}\tabularnewline
\textbf{Name} & LIF neuron\tabularnewline
\textbf{Type} & leaky integrate-and-fire, exponential synaptic current inputs\tabularnewline
\textbf{Subthreshold dynamics} & $\frac{\d V}{\d t}=-\frac{V-\EL}{\tau_{\mathrm{m}}}+\frac{\Is(t)}{C_{\mathrm{m}}}$\hspace*{0.1cm}
if $\mathrm{\left(t>t^{*}+\taur\right)}$\newline $V(t)=\Vr$\hspace*{1cm}
else 

$\Is(t)=\sum_{i,k}J_{k}\,e^{-(t-t_{i}^{k})/\taus}\Theta(t-t_{i}^{k})$
$k$: neuron index, $i$: spike index\tabularnewline
\textbf{Spiking} & If $V(t-)<\theta\wedge V(t+)\geq\theta$\newline1. set $t^{*}=t$,
2. emit spike with time stamp $t^{*}$\tabularnewline
\end{tabular}

\begin{tabular}{@{\hspace*{1mm}}p{3cm}@{}@{\hspace*{1mm}}p{3cm}@{}@{\hspace*{1mm}}p{9.1cm}}
\multicolumn{3}{>{\color{white}\columncolor{black}}c}{\textbf{E: Input}}\tabularnewline
\textbf{Type} & \textbf{Target} & \textbf{Description}\tabularnewline
Background & LIF neurons & independent Poisson spikes (see \prettyref{suptab:Simulation-parameters})\tabularnewline
\end{tabular}

\begin{tabular}{l||c}
\multicolumn{2}{>{\color{white}\columncolor{black}}c}{\textbf{F: Measurements}}\tabularnewline
\hline 
\multicolumn{2}{p{15.2cm}}{Spiking activity}\tabularnewline
\end{tabular}

\caption{Model description after \citet{Nordlie-2009_e1000456}.}
\label{tab:Model-description}
\end{table}

In the following, we detail how we derive the structure of the model
(summarized in \prettyref{tab:Model-description}), i.e., the population
sizes, the local and cortico-cortical connectivity and the external
drive.

\subsection{Numbers of neurons}

We estimate the number of neurons $N(A,i)$ in population $i$ of
area $A$ in three steps:
\begin{enumerate}
\item We translate neuronal volume densities to the FV91 scheme from the
most representative area in the original scheme (\prettyref{suptab:translate_arch_types}).
For areas not covered by the data set, we take the average laminar
densities for areas of the same architectural type. Table 4 of \citet{Hilgetag15_submitted}
lists the architectural types, which we translate to the FV91 scheme
according to \prettyref{suptab:translate_arch_types}. To the previously
unclassified areas MIP and MDP we manually assign type 5 like their
neighboring area PO, which is similarly involved in visual reaching
\citep{Johnson96_102,Galletti03_158}, and was placed at the same
hierarchical level by \citet{Felleman91_1}.  
\item We determine total and laminar thicknesses as detailed in \nameref{sec:results}. 
\item The fraction $\gamma(v)$ of excitatory neurons in layer $v$ is taken
to be identical across areas. For the laminar dependency, values from
cat V1 \citep{Binzegger04} are used with 78\% excitatory neurons
in layer 2/3, 80\% in L4, 82\% in L5, and 83\% in L6.
\end{enumerate}
The resulting number of neurons in population $i$ of area $A$ is

\begin{eqnarray}
N(A,i)=\rho\left(A,v{}_{i}\right)S(A)D\left(A,v{}_{i}\right)\cdot\negthickspace\negthickspace\negthickspace & \begin{cases}
\gamma\left(v{}_{i}\right) & \textrm{if }i\in\mathcal{E}\\
1-\gamma\left(v{}_{i}\right) & \textrm{\text{if }}i\in\mathcal{I}
\end{cases} & ,\label{eq:neuronnumber}
\end{eqnarray}
where $v_{i}$ denotes the layer of population $i$, $S(A)$ the surface
area of area $A$ (cf.\,\prettyref{suptab:Area_surfaces}), $D(A,v{}_{i})$
the thickness of layer $v{}_{i}$, and $\mathcal{E},\mathcal{\,I}$
the pool of excitatory and inhibitory populations, respectively. \prettyref{suptab:neuron_numbers}
gives the population sizes corresponding to the modeled $1\mmsquare$
area size.

\subsection{Local connectivity\label{sub:Local-connectivity}}

The connection probabilities of the microcircuit model \citep[Table 5]{Potjans14_785},
computed from anatomical and electrophysiological studies \citep[with large contributions from][]{Binzegger04,Thomson07_19},
form the basis for the local circuit of each area. The connectivity
between any pair of populations is spatially uniform. However, we
take the underlying probability $C$ for a given neuron pair to establish
one or more contacts to decay with distance according to a Gaussian
with standard deviation $\sigma=297\,\mathrm{\mu m}$ \citep{Potjans14_785}.
We approximate each brain area as a flat disk with (area-specific)
radius $R$ and assign polar coordinates $r$ and $\theta$ to each
neuron. The radius determines the cut-off of the Gaussian and hence
the precise connectivities. The average connection probability is
obtained by integrating over all possible positions of the two neurons: 

\begin{equation}
\bar{C}(R)=\frac{C_{0}}{\pi^{2}R^{4}}\int_{0}^{R}\hspace{-6pt}\int_{0}^{2\pi}\hspace{-8pt}\int_{0}^{R}\hspace{-6pt}\int_{0}^{2\pi}\exp\left[\frac{-\left(r_{1}^{2}+r_{2}^{2}-2r_{1}r_{2}\cos(\theta_{1}-\theta_{2})\right)}{2\sigma^{2}}\right]r_{1}r_{2}\d\theta_{1}\d r_{1}\d\theta_{2}\d r_{2}\,,\label{eq:mean_prob}
\end{equation}
with $C_{0}$ the connection probability at zero distance. This can
be reduced to a simpler form \citep{Sheng85},

\begin{equation}
\bar{C}(R)=\frac{2C_{0}}{\pi R^{2}}\int_{0}^{2R}\hspace{-5pt}e^{-r^{2}/2\sigma^{2}}\hspace{-4pt}\left[4\arctan\hspace{-2pt}\left(\frac{2R-r}{2R+r}\right)^{1/2}\hspace{-4pt}-\sin\hspace{-2pt}\left(4\arctan\left[\frac{2R-r}{2R+r}\right]^{1/2}\right)\right]r\d r\,.\label{eq:sheng}
\end{equation}
Averaged across population pairs, $C_{0}$ is $0.143$ \citep[computed from Eq.\ 8 and Table S1 in][]{Potjans14_785}.
Note that \citet{Potjans14_785} only vary the position of one neuron,
keeping the other neuron fixed in the center of the disk (Eq.\,9
in that paper). Henceforth, we denote connection probabilities computed
with the latter approach with the subscript $\mathrm{PD14}$ and use
primes for all variables referring to a network with the population
sizes of the microcircuit model.

The parameters of the microcircuit model are reported for a $1\mmsquare$
patch of cortex, corresponding to \textrm{$R=\sqrt{1/\pi}\,\mathrm{mm}$},
which we call $R_{0}$. For each source population $j$ and target
population $i$, we first translate the connection probabilities of
the \textrm{$1\mmsquare$} model to area-dependent $R$ via

\[
C_{ij}^{\prime}(R)=C_{ij,\mathrm{PD14}}^{\mathrm{\prime}}\left(R_{0}\right)\frac{\bar{C}^{\prime}(R)}{\bar{C}_{\mathrm{PD14}}^{\prime}\left(R_{0}\right)}\,,
\]
with $\bar{C}_{\mathrm{PD14}}^{\prime}(R_{0})=0.066$. From this,
we compute the number of synapses
\[
N_{\mathrm{syn},ij}=\frac{\log\left(1-C_{ij}\right)}{\log\left(1-\frac{1}{N_{i}N_{j}}\right)}\,,
\]
based on randomly drawing source and target neurons with replacement
\citep[cf.\ Eq.\ 1 in][]{Potjans14_785}. The indegree $K_{ij}$ is
the number of incoming synapses per target neuron, $N_{\mathrm{syn},ij}/N_{i}$.
Henceforth, all numbers of synapses $N_{\mathrm{syn}}(A)$ and indegrees
$K_{ij}(A)$ are area-specific. For simplicity, we drop the argument
$A$. Since mean synaptic inputs are proportional to the indegrees,
we consider them a defining characteristic of the local circuit and
preserve their relative values when adjusting the model to area-specific
population sizes,

\[
\frac{K_{ij}(R)}{K_{kl}(R)}=\frac{K_{ij}^{\prime}(R)}{K_{kl}^{\prime}(R)}\,\,\,\forall i,j,k,l
\]
\begin{equation}
\Leftrightarrow K_{ij}(R)=c_{A}(R)K_{ij}^{\mathrm{\prime}}(R)\,\,\,\forall i,j\,,\label{eq:indegree_equation}
\end{equation}
with $c_{A}(R)$ an area-specific conversion factor, which is larger
for areas with smaller neuron densities because of the assumption
of constant synaptic volume density. It is computed as 
\[
c_{A}(R)=\frac{N_{\mathrm{syn,tot}}(R)}{\sum_{i,j}N_{i}K_{ij}^{\prime}}\,\FLN_{\mathrm{i}}\Bigg\langle\frac{K_{ij}^{\prime}(R)}{K_{ij}^{\prime}(R_{\mathrm{full}})}\Bigg\rangle_{ij}\,,
\]
with $FLN_{\mathrm{i}}$ the fraction of labeled neurons intrinsic
to the injected area in a retrograde tracing experiment by \citet{Markov11_1254}
and $N_{\mathrm{syn,tot}}(R)=\rho_{\mathrm{syn}}\pi R^{2}D$ with
$D$ the total thickness of the given area. For details, see \nameref{sec:Supplement-Exp}.

\subsection{Cortico-cortical connectivity\label{sub:Cortico-cortical-connectivity}}

We determine whether a pair of areas is connected using the union
of all connections reported in the FV91 scheme in the CoCoMac database
\citep{Stephan01_1159,Bakker12_30,Suzuki94_1856,Felleman91_1,Rockland79_3,Barnes92_222}
(\prettyref{fig:construction}F, see \nameref{sec:Supplement-Exp}
for details) and all connections reported by \citet{Markov2014_17}.
We then determine population-specific numbers of modeled cortico-cortical
synapses in three steps: 1.\,\,deriving the area-level connectivity;
2. distributing synapses across layers; 3. assigning synapses to target
neurons. 

For the first step, we compute the total number of synapses formed
between each pair of areas using retrograde tracing data from \citet{Markov2014_17}.
The data consist of fractions of labeled neurons $\FLN_{AB}=NLN_{AB}/\sum_{B^{\prime}}NLN_{AB^{\prime}}$,
with $NLN_{AB}$ the number of labeled neurons in area $B$ upon injection
in area $A$. \citet{Markov2014_17} used a parcellation scheme called
M132 which is also available as a cortical surface, both in native
and in F99 space. On the target side we use the coordinates of the
injection sites registered to the F99 atlas available via the Scalable
Brain Atlas \citep{Bakker15} to identify the equivalent area in the
FV91 parcellation (cf.\,\prettyref{suptab:Injected-areas-markov}).
There is data for 11 visual areas in the FV91 scheme with repeat injections
in six areas, for which we take the arithmetic mean. To map data on
the source side from M132 to FV91, we count the number of overlapping
triangles on the F99 surface between any given pair of regions and
distribute the $FLN$ proportionally to the amount of overlap, using
the F99 region overlap tool at the CoCoMac site (http://cocomac.g-node.org).
To estimate values for the areas not included in the data set, we
use an exponential decay of connectivities with distance \citep{ErcseyRavasz13},
\begin{equation}
FLN_{AB}=C\cdot\exp\left(-\lambda d_{AB}\right)\,.\label{eq:EDR}
\end{equation}
A linear least-squares fit of the logarithm of the $FLN$ (\prettyref{fig:construction}G)
predicts missing values. The total number of synapses $N_{\mathrm{syn},AB}$
between each pair of areas $\{A,B\}$ is assumed to be proportional
to the number of labeled neurons $NLN_{AB}$ and thus to $FLN_{AB}$,

\[
\frac{N_{\mathrm{syn},AB}}{\underbrace{\sum_{B^{\prime}}N_{\mathrm{syn,}AB'}}_{=N_{\mathrm{syn,tot},A}}}=\frac{NLN_{AB}}{\sum_{B^{\prime}}NLN_{AB^{\prime}}}=\frac{FLN_{AB}}{\sum_{B'}FLN_{AB'}}.
\]
This corresponds to individual neurons in each source area (including
area $A$ itself) on average establishing the same number of synapses
in the target area $A$. For each target area, the $FLN$ in the model
should add up to the total fraction of connections from visual cortical
areas, which is not known a priori. For normalization, we consider
also non-visual areas, for which distances are available and for which
we can hence also estimate the $FLN$. The total fraction of all
connections from subcortical regions averages $1.3\,\%$ in eight
cortical areas \citep{Markov11_1254}. This allows us to normalize
the combined $FLN$ from all cortical areas as $\sum_{B}FLN_{AB}=1-FLN_{\mathrm{i}}-0.013$,
where the sum includes both modeled and non-modeled cortical areas.

As a next step, we determine the distribution of connections across
source and target layers.  On the source side, the laminar projection
pattern can be expressed as the fraction of supragranular labeled
neurons ($\SLN$) in retrograde tracing experiments \citep{Markov14}.
To determine the $SLN$ entering into the model, we use the exact
coordinates of the injections to determine the corresponding target
area $A$ in the FV91 parcellation, and for each pair of areas we
take the mean $SLN$ across injections. To map the data from M132
to FV91, we weight the $SLN$ by the overlap $c_{B,\beta}$ between
area $\beta$ in the former and area $B$ in the latter scheme and
the $FLN$ to take into account the overall strength of the connection,
\[
SLN_{AB}=\frac{\sum_{\beta}c_{B,\beta}FLN_{A,\beta}SLN_{A,\beta}}{\sum_{\beta}c_{B,\beta}FLN_{A,\beta}}\,.
\]
We estimate missing values using a sigmoidal fit of $SLN$ vs. the
logarithmized ratio of overall cell densities of the two areas (\prettyref{fig:construction}J).
A relationship between laminar patterns and log ratios of neuron densities
was suggested by \citet{Beul15_arxiv}. Following \citet{Markov14},
we use a generalized linear model and assume the numbers of labeled
neurons in the source areas to sample from a beta-binomial distribution
\citep[e.g.][]{Weisstein05}. This distribution arises as a combination
of a binomial distribution with probability $p$ of supragranular
labeling in a given area, and a beta distribution of $p$ across areas
with dispersion parameter $\phi$. With the probit link function $g$
\citep[e.g.][]{McCulloch08}, the measured $SLN$ relates to the log
ratio $\ell$ of neuron densities for each pair of areas as

\begin{equation}
g(SLN)=a_{0}\left(\begin{array}{c}
1\\
\vdots\\
1
\end{array}\right)+a_{1}\ell,\label{eq:SLN_GLM}
\end{equation}
where $\ell$ and $SLN$ are vectors and $\{a_{0},a_{1}\}$ are scalar
fit parameters. To fit $SLN$ vs. log ratios of cell densities, we
map the FV91 areas to the \citet{Markov14} scheme with the overlap
tool of CoCoMac (see above) and compute the cell density of each area
in the M132 scheme as a weighted average over the relevant FV91 areas.
For areas with identical names in both schemes, we simply take the
neuron density from the FV91 scheme. \prettyref{fig:construction}J
shows the result of the $SLN$ fit in R \citep{RSoftware} with the
\texttt{betabin} function of the aod package \citep{Lesnoff12}. In
contrast to \citet{Markov14}, who exclude certain areas when fitting
$SLN$ vs. hierarchical distances in view of ambiguous hierarchical
relations, we take all data points into account to obtain a simple
and uniform rule.

As a further step, we combine $SLN$ with CoCoMac data. The data sets
complement each other: $SLN$ provides quantitative data on laminar
patterns of incoming projections for about one quarter of the connected
areas.  CoCoMac has values for all six layers, but limited to a
qualitative strength ranging from 0 (absent) to 3 (strong) which we
take to represent numbers of synapses in orders of magnitude (see
\nameref{sec:Supplement-Exp}). Whether or not to include a layer
in source pattern $P_{\mathrm{s}}$ is based on CoCoMac \citep{Felleman91_1,Barnes92_222,Suzuki94_497,Morel90_555,Perkel86,Seltzer94}
if the corresponding data is available ($45\,\%$ coverage); otherwise,
we include L2/3, L5 and L6 and exclude L4 \citep{Felleman91_1}. We
model cortico-cortical connections as purely excitatory, a good approximation
to experimental findings \citep{Salin95_107,Tomioka07_526}. If a
layer is included in the source pattern, we assign a fraction of the
total outgoing synapses to it according to the $SLN$. Since the $SLN$
do not further distinguish between the infragranular layers 5 and
6, we use the rough connection densities from CoCoMac for this purpose
when available, and otherwise we distribute synapses in proportion
to the numbers of neurons. On the target side, we determine the pattern
of target layers $P_{\mathrm{t}}$ from anterograde tracer studies
in CoCoMac \citep{Jones78_291,Rockland79_3,Morel90_555,Webster91_1095,Felleman91_1,Barnes92_222,Distler93_125,Suzuki94_497,Webster94_470}
if available (29\% coverage); otherwise we use termination patterns
suggested by the $SLN$ based on a relationship between source and
target patterns. Using the terminology of visual connection hierarchies,
we denote projections with low, intermediate, and high $SLN$ respectively
as feedback, lateral, and feedforward projections. We take $SLN<0.35$
to correspond to feedback projections, $SLN>0.65$ to feedforward
projections and $SLN\in[0.35,0.65]$ to lateral projections. The corresponding
termination patterns $P_{\mathrm{t}}$ are

\[
\begin{split}\{4\}\,\text{for }SLN>0.65\\
\{1,2/3,5,6\}\,\text{for }SLN<0.35\\
\{1,2/3,4,5,6\}\,\text{for }SLN\in[0.35,0.65]
\end{split}
\,,
\]
and we distribute synapses among the layers in the termination pattern
in proportion to their thickness.

Since we use a point neuron model, we have to account for the possibly
different laminar positions of cell bodies and synapses. The data
of \citet{Binzegger04} deliver three quantities that allow us to
relate synapse to cell body locations: first, the probability $\mathcal{P}(s_{\mathrm{cc}}|c_{\mathrm{B}}\bigcap s\in v)$
for a synapse in layer $v$ on a cell of type $c_{\mathrm{B}}$ (e.g.,
a pyramidal cell with soma in L5) to be of cortico-cortical origin;
second, the relative occurrence $\mathcal{\mathcal{\mathcal{P}}}(c_{\mathrm{B}})$
of the cell type $c_{\mathrm{B}}$; and third, the total numbers of
synapses $N_{\mathrm{syn}}(v,c_{\mathrm{B}})$ in layer $v$ onto
the given cell type. We map these data to our model by computing the
conditional probability $\mathcal{P}(i|s_{\mathrm{cc}}\in v)$ for
the target neuron to belong to population $i$ if a cortico-cortical
synapse $s_{\mathrm{cc}}$ is in layer $v$. This probability equals
the sum of probabilities that a synapse is established on the different
\citeauthor{Binzegger04} subpopulations making up our populations,

\begin{equation}
\mathcal{P}(i|s_{\mathrm{cc}}\in v)=\mathcal{P}(\bigcup_{\mathrm{c_{\mathrm{B}}}\in i}c_{\mathrm{B}}|s_{\mathrm{cc}}\in v)=\sum_{c_{\mathrm{B}}\in i}\mathcal{P}(c_{\mathrm{B}}|s_{\mathrm{cc}}\in v)\,,
\end{equation}
where

\begin{equation}
\mathcal{P}(c_{\mathrm{B}}|s_{\mathrm{cc}}\in v)=\frac{\mathcal{P}(c_{\mathrm{B}}\bigcap s_{\mathrm{cc}}\in v)}{\mathcal{P}(s_{\mathrm{cc}}\in v)}\,.\label{eq:synapse_to_cell_body}
\end{equation}
The numerator gives the joint probability that a cortico-cortical
synapse is formed in layer $v$ on cell type $c_{\mathrm{B}}$,
\[
\mathcal{P}(c_{\mathrm{B}}\bigcap s_{\mathrm{cc}}\in v)=\frac{N_{\mathrm{syn,CC}}(v,c_{\mathrm{B}})\mathcal{P}(c_{\mathrm{B}})}{\sum_{v^{\prime},c_{\mathrm{B}}^{\prime}}N_{\mathrm{syn,CC}}(v{}^{\prime},c_{\mathrm{B}}^{\prime})\mathcal{P}(c_{\mathrm{B}}^{\prime})}\,,
\]
and the denominator is the probability of a cortico-cortical synapse
in layer $v$, computed by summing over cell types,

\[
\mathcal{P}(s_{\mathrm{cc}}\in v)=\sum_{c_{\mathrm{B}}}\mathcal{P}(c_{\mathrm{B}}\bigcap s_{\mathrm{cc}}\in v)\,.
\]
$N_{\mathrm{syn,CC}}(v,c_{\mathrm{B}})$ represents the number of
cortico-cortical synapses in layer $v$ on cell type $c_{\mathrm{B}}$,

\[
N_{\mathrm{syn,CC}}(v,c_{\mathrm{B}})=\mathcal{P}(s_{\mathrm{cc}}|c_{\mathrm{B}}\bigcap s\in v)\,N_{\mathrm{syn}}(v,c_{\mathrm{B}})\,\mathcal{\mathcal{\mathcal{P}}}(c_{\mathrm{B}})\,,
\]
which can be directly determined from the data. Combining these equations,
we obtain the number of cortico-cortical (type III) synapses from
excitatory population $j$ of area $B$ to population $i$ of area
$A$ (cf.\,\prettyref{fig:construction}K):

\begin{eqnarray}
N_{\mathrm{syn},\mathrm{III}}(i,A,j,B) & = & \underbrace{Z_{i}\sum_{v\in P_{\mathrm{t}}}Y_{v}\,\mathcal{P}(i|s_{\mathrm{cc}}\in v)}_{\text{target side}}\,\underbrace{X_{j}}_{\text{source side}}N_{\mathrm{syn},\mathrm{III}}(A,B)\,,\label{eq:cc_syn_distribution}\\
\text{with \,}X_{j} & = & \begin{cases}
\SLN & \textrm{if }j\in S\bigcap P_{\text{s}}\,\\
(1-\SLN)\frac{10^{\alpha(v_{j})}}{\sum_{j^{\prime}\in I,\,\alpha(v_{j^{\prime}})>0}10^{\alpha(v_{j^{\prime}})}} & \text{if }j\in I\,\text{and }\alpha(v_{j})>0\\
(1-\SLN)\frac{N(A,j)}{_{\sum_{j'\in I}N(A,j^{\prime})}} & \textrm{\text{if }}j\in I\bigcap P_{\text{s}}\text{\, but no CoCoMac data available}\\
0 & \text{if }j\notin P_{\mathrm{s}}
\end{cases}\,,\nonumber \\
\text{and \,}Y_{v} & = & \begin{cases}
\frac{10^{\alpha(v)}}{\sum_{\alpha(v^{\prime})>0}10^{\alpha(v^{\prime})}} & \text{if\,}\alpha(v)>0\\
\frac{D(A,v)}{\sum_{v^{\prime}}D(A,v^{\prime})} & \text{if no CoCoMac data available\,}
\end{cases}\,.\nonumber 
\end{eqnarray}
Here, $S=\text{2/3E}$ and $I=\left\{ \text{5E,6E}\right\} $ respectively
denote the supragranular and infragranular excitatory populations.
$Z_{i}$ is an additional factor which takes into account that cortico-cortical
feedback connections preferentially target excitatory rather than
inhibitory neurons \citep{Johnson96_383,Anderson11}. $Z_{i}$ is
area-specific and depends on the excitatory or inhibitory nature of
the target population, but not on the target layer. We choose a fraction
of $93\,\%$ of connections targeting excitatory neurons, as an average
over experimental values ranging between $87\,\%$ and $98\,\%$.
For each feedback connection in the model, we thus redistribute the
synapses across the excitatory and inhibitory target populations and
determine $Z_{i}$ such that 
\[
\frac{\sum_{i\in\mathcal{E}}\sum_{j}N_{\mathrm{syn},\mathrm{III}}(i,A,j,B)}{N_{\mathrm{syn},\mathrm{III}}(A,B)}=0.93\,.
\]
\prettyref{supfig:connectivity-pops} shows the resulting connection
probabilities between all population pairs in the model.

\subsection{External, random input}

Since quantitative area-specific data on non-visual and subcortical
inputs are highly incomplete, we use a simple scheme to determine
numbers of external inputs: For each area, we compute the total number
of external synapses as the difference between the total number of
synapses and those of type I and III and distribute these such that
all neurons in the given area have the same indegree for Poisson sources.
In area TH, we compensate for the missing granular layer 4 by increasing
the external drive onto populations 2/3E and 5E by $20\,\%$. With
the modified connectivity matrix yielded by the analytical procedure
described in \citet{Schuecker15_arxiv}, we set $\kappa=1.125$ to
increase the external indegree onto population 5E by $12.5\,\%$ and
onto 6E by $42\,\%$ to elevate the firing rates in these populations.
\prettyref{suptab:external-input} lists the resulting external indegrees.

\subsection{Network simulations}

We performed simulations on the JUQUEEN supercomputer \citep{stephan2015juqueen}
with NEST version 2.8.0 \citep{Nest280} with optimizations for the
use on the supercomputer which will be included in a future release.
All simulations use a time step of $0.1\ms$ and exact integration
for the subthreshold dynamics of the LIF neuron model \citep[reviewed in][]{Plesser09_353}.
Simulations were run for $100.5\s$ ($\lambda=1.9$), $50.5\s$ ($\lambda\in[1.8,\,2.0,\,2.1]$),
and $10.5\ms$ ($\lambda\in[1.,\,1.5,\,1.7,\,2.5]$) biological time
discarding the first $500\ms$. Spike times were recorded from all
neurons, except for the simulations shown in \prettyref{fig:bistability}A,B,
where we recorded from 1000 neurons per population.

\subsection{Analysis methods}

We investigate the structural properties of the model with the map
equation method \citep{Rosvall10}. In this clustering algorithm,
an agent performs random walks between graph nodes with probability
proportional to the outdegree of the present node and a probability
($p=0.15$) of jumping to a random network node. The algorithm detects
clusters in the graph by minimizing the length of a binary description
of the network using a Huffman code. To assess the quality of the
clustering, we use a modularity measure which extends a measure for
unweighted, directed networks \citep{Leicht08} to weighted networks,
analogous to \citealt{Newman04}, 
\[
Q=\frac{1}{m}\sum_{A,B}\left(\mathcal{O}_{AB}-\frac{\sum_{B^{\prime}}\mathcal{O}_{AB^{\prime}}\cdot\sum_{A^{\prime}}K_{A^{\prime}B}}{m}\right)\delta_{\mathcal{C}_{A},\mathcal{C}_{B}}\,,
\]
where $\mathcal{O}_{AB}$($K_{AB}$) is the matrix of relative outdegrees
(indegrees), $m=\sum_{A,B}\mathcal{O}_{A,B}$ and $\delta_{\mathcal{C}_{A},\mathcal{C}_{B}}=1$
if areas $A$ and $B$ are in the same cluster and 0 otherwise. $Q=0$
reflects equal connectivity within and between clusters, while $Q=1$
corresponds to connectivity exclusively within clusters.

Instantaneous firing rates are determined as spike histograms with
bin width $1\ms$ averaged over the entire population or area. In
\prettyref{fig:ground_state}G, \prettyref{supfig:ground_state_lambda1}G,
and to determine the temporal hierarchy, we convolve the histograms
with Gaussian kernels with $\sigma=2\ms$. Spike-train irregularity
is quantified for each population by the revised local variation $LvR$
\citep{Shinomoto09_e433} averaged over a subsample of $2000$ neurons.
The cross-correlation coefficient is computed with bin width $1\ms$
on single-cell spike histograms of a subsample of $2000$ neurons
per population with at least one emitted spike per neuron. 

The single-cell autocorrelation function is calculated on spike histograms
with bin width $2.5\ms$ to suppress fast fluctuations on the order
of the refractory time, normalized to the zero-lag peak, and averaged
across a subsample of $2000$ neurons. We then perform a linear least-squares
fit $f(t)=A-t/\tau$ on the logarithmized autocorrelation for all
times $t\in\left[2.5,\,75\right]$ and define the inverse slope $\tau$
as the intrinsic time scale of the population. If the autocorrelation
drops to a local minimum at the first time lag $t=2.5\ms$, we set
the time scale to the refractory period, $\tau=2\ms$.

The temporal hierarchy is based on the cross-covariance function between
area-averaged firing rates. We use a wavelet-smoothing algorithm (\texttt{signal.find\_peaks\_cwt}
of python scipy library \citep{scipy01} with peak width $\Delta=20\ms$)
to detect extrema for $\tau\in[-100,\,100]$ and take the location
of the extremum with the largest absolute value as the time lag. 

Functional connectivity (FC) is defined as the zero-time lag cross-correlation
coefficient of the area-averaged synaptic inputs 
\[
I_{A}(t)=\frac{1}{N_{A}}\sum_{i\in A}N_{i}\left|I_{i}(t)\right|=\frac{1}{N_{A}}\sum_{i\in A}N_{i}\sum_{j}K_{ij}\left|J_{ij}\right|\left(\nu_{j}\ast PSC_{j}\right)(t)\:,
\]
 with the normalized post-synaptic current $PSC_{j}(t)=\exp[-t/\tau_{\mathrm{s}}]$,
the population firing rate $\nu_{j}$ of source population $j$, indegree
$K_{ij}$, and synaptic weight $J_{ij}$ of the connection from $j$
to target population $i$ containing $N_{i}$ neurons.

The clustering of the FC matrices was performed using the function
\texttt{modularity\_louvain\_und\_sign} of the Brain Connectivity
Toolbox (BCT; http://www.brain-connectivity-toolbox.net) with the
$Q^{\ast}$ option, which weights positive weights more strongly than
negative weights, as introduced by \citet{Rubinov11_2068}.

\subsection{Macaque resting-state fMRI}

Data were acquired from six male macaque monkeys (4 \textit{Macaca
mulatta} and 2 \textit{Macaca fascicularis}). All experimental protocols
were approved by the Animal Use Subcomittee of the University of Western
Ontario Council on Animal Care and in accordance with the guidelines
of the Canadian Council on Animal Care. Data acquisition, image preprocessing
and a subset of subjects (5 of 6) were previously described \citep{Babapoor2013}.
Briefly, 10 5-min resting-state fMRI scans (TR: $2\s$; voxel size:
$1\mm$ isotropic) were acquired from each subject under light anaesthesia
($1.5
$ isoflurane). Additional processing for the current study included
the regression of nuisance variables using the AFNI software package
(\url{afni.nimh.nih.gov/afni}), which included six motion parameters
as well as the global white matter and CSF signals. The global mean
signal was not regressed.

The FV91 parcellation was drawn on the F99 macaque standard cortical
surface template \citep{VanEssen01_443} and transformed to volumetric
space with a $2\mm$ extrusion using the Caret software package (\url{http://www.nitrc.org/projects/caret}).
The parcellation was applied to the fMRI data and functional connectivity
computed as the Pearson correlation coefficients between probabilistically-weighted
ROI timeseries for each scan \citep{Shen12}. Correlation coefficients
were Fisher z-transformed and correlation matrices were averaged within
animals and then across animals before transforming back to Pearson
coefficients.

%% file: supplement.tex
\section{Supplemental Figures}

\begin{center}
\begin{figure}[H]
\begin{centering}
\includegraphics{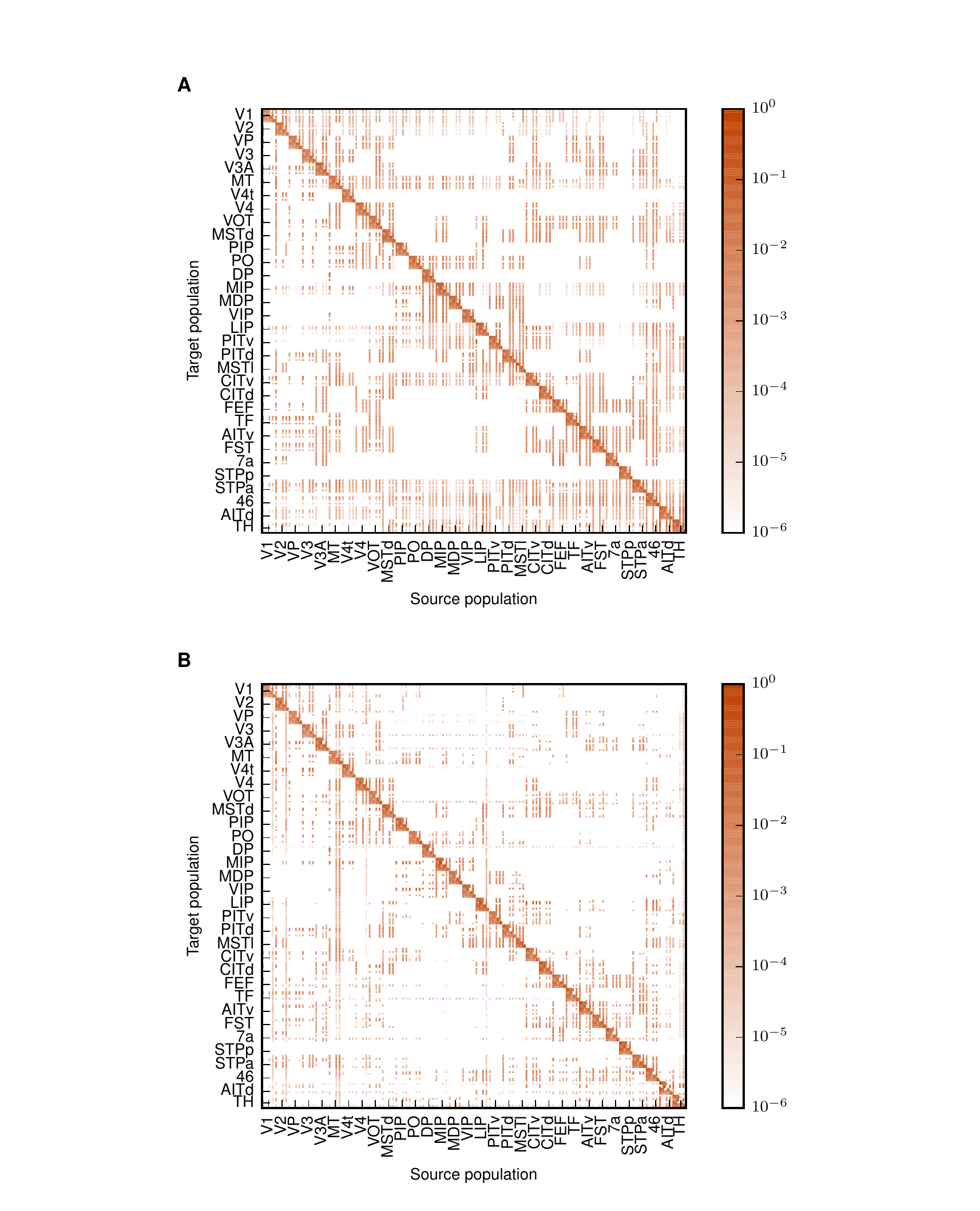}
\par\end{centering}

\caption{Related to \prettyref{fig:construction}. \textbf{Connection probabilities
of the model} encoded in color before (A) and after (B) applying the
theoretical method described in \citet{Schuecker15_arxiv}. Areas
are ordered according to their architectural types, and populations
inside the areas are ordered as {[}2/3E, 2/3I, 4E/, 4I, 5E, 5I, 6E,
6I{]}.}
\label{supfig:connectivity-pops}
\end{figure}

\par\end{center}

\begin{center}
\begin{figure}[H]
\includegraphics{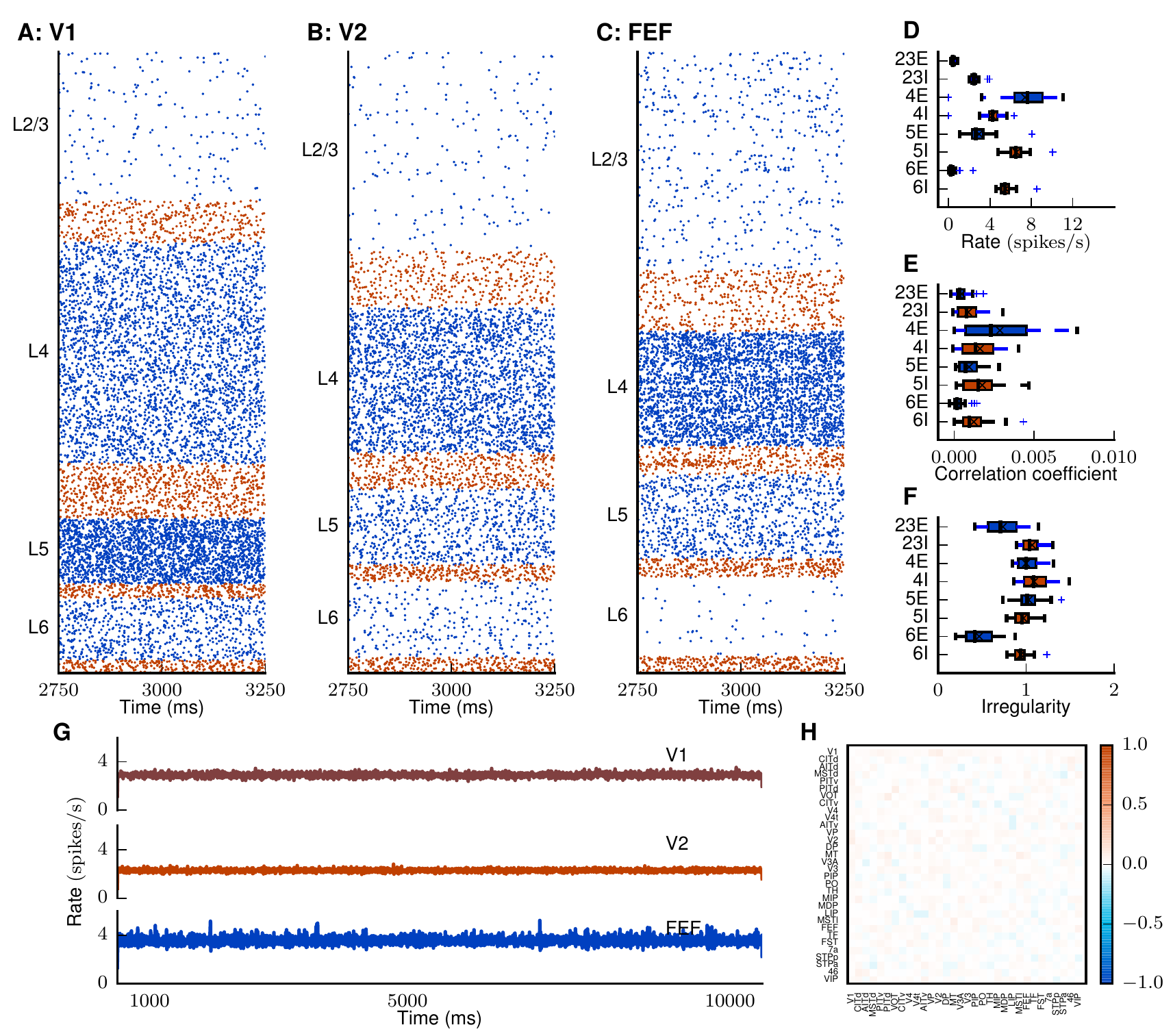}

\caption{Related to \prettyref{fig:ground_state}. \textbf{Resting state of
the model with cortico-cortical synaptic weights equal to intrinsic
synaptic weights.} (A-C) Raster plot of spiking activity of $5\,\%$
of the neurons in area V1 (A), V2 (B), and FEF (C). (D-F) Statistics
of the spiking activity across areas and populations shown as area-averaged
box plots \citep{Tukey77}. (D) Population-averaged firing rates.
The firing rates differ across areas and layers, varying between $0.05$
and $25\protect\spikess$. Inhibitory populations are more active
than excitatory populations across layers and areas (with the exception
of layer 5 in V1 and layer 4 in higher areas) despite the identical
intrinsic properties of the two cell types. The excitatory populations
of layer 2/3 and 6 exhibit lower firing rates than those of layers
4 and 5, similar to the microcircuit model \citep{Potjans14_785}
(E) Average pairwise cross-correlation coefficients of spiking activity.
(F) Irregularity measured by revised local variation $LvR$ \citep{Shinomoto09_e433}
averaged across neurons. (G) Area-averaged firing rates. Parameters
are $g=-11,\:\nu_{\mathrm{ext}}=10\protect\spikess,\:\kappa=1.125,\:\lambda=1,\:\lambda_{\mathcal{I}}=1$.
(H) Functional connectivity (FC) in the model for measured by the
zero-time lag cross-correlation coefficiens. Functional connectivity
between areas is very low for most pairs of areas. Areas are ordered
as in \prettyref{fig:interactions}A.}
\label{supfig:ground_state_lambda1}
\end{figure}

\par\end{center}

\begin{center}
\begin{figure}[H]
\includegraphics{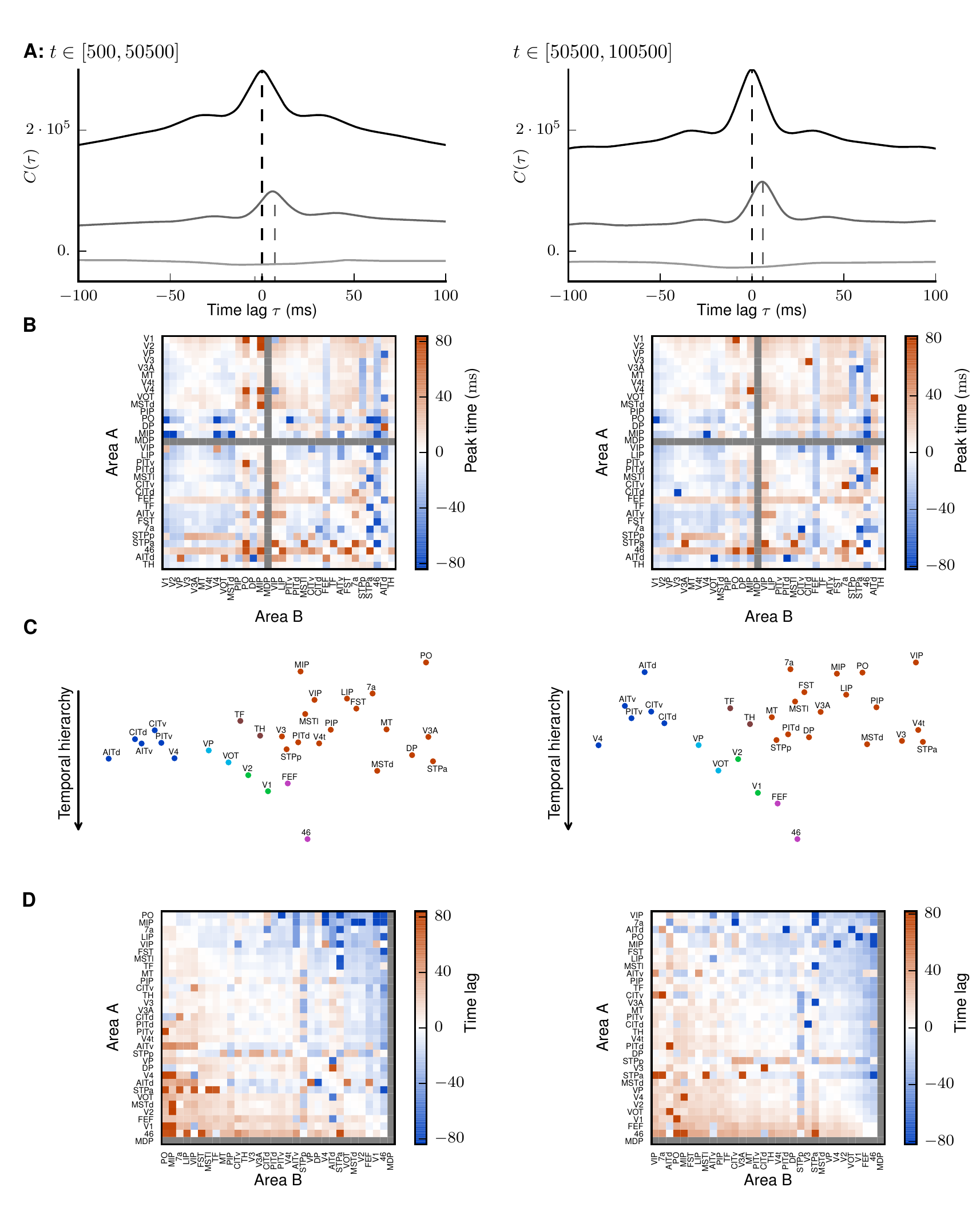}

\caption{Related to \prettyref{fig:correlation_hierarchy}. \textbf{Validation
of the temporal hierarchy. }Left column: First half of the time series
($t\in[500\protect\ms,\,50,500\protect\ms]$), right column: second
half of the time series ($t\in[50,500\protect\ms,\,100,500\protect\ms]$).
(A) Covariance functions of the area-averaged firing rates of V1 with
areas V2 (gray) and FEF (light gray), and auto-covariance function
of V1 (black). Dashed lines mark peak positions, detected by a wavelet
smoothing algorithm (see \nameref{sec:exp_procedures}). (B) Matrix
of peak positions of the correlation function for all pairs of areas.
Area MDP was neglected because it has only outgoing but no incoming
connections to other visual areas according to CoCoMac. (C) Temporal
hierarchy. Colors indicate the cluster of each area found with the
map equation algorithm (cf. \prettyref{fig:map_equation}). Areas
are horizontally arranged to avoid visual overlap. (D) Peak position
matrix with areas in hierarchical order.}
\label{supfig:validation_hierarchy}
\end{figure}
\begin{figure}[t]
\includegraphics{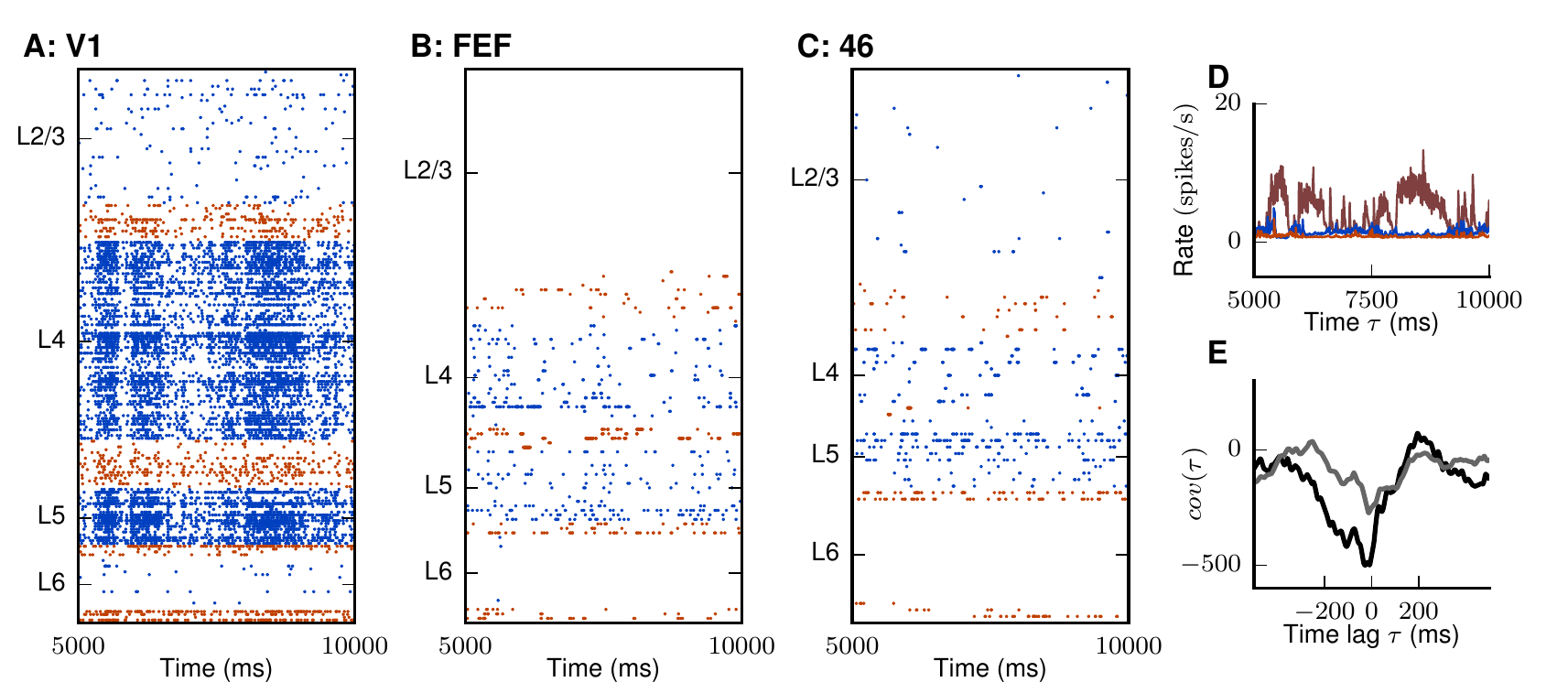}\caption{Related to \prettyref{fig:interactions}. \textbf{Anticorrelation
between V1 and frontal areas.} (A-C) Raster plot of spiking activity
of $0.1\,\%$ of the neurons in area V1 (A), FEF (B), and 46 (C).
(D) Area-averaged firing rates of V1 (brown), FEF (red) and 46 (blue).
(E) Covariance functions of the area-averaged firing rates of V1 with
areas 46 (black) and FEF (gray). Parameters are the same as for the
simulation used in \prettyref{fig:interactions}. The plot shows the
anticorrelation between V1 and the two frontal areas visible in both
the raster and the rate plot as well as in the negatively peaked cross-covariance
function.}
\label{supfig:anticorrelation}
\end{figure}
\begin{figure}[t]

\includegraphics{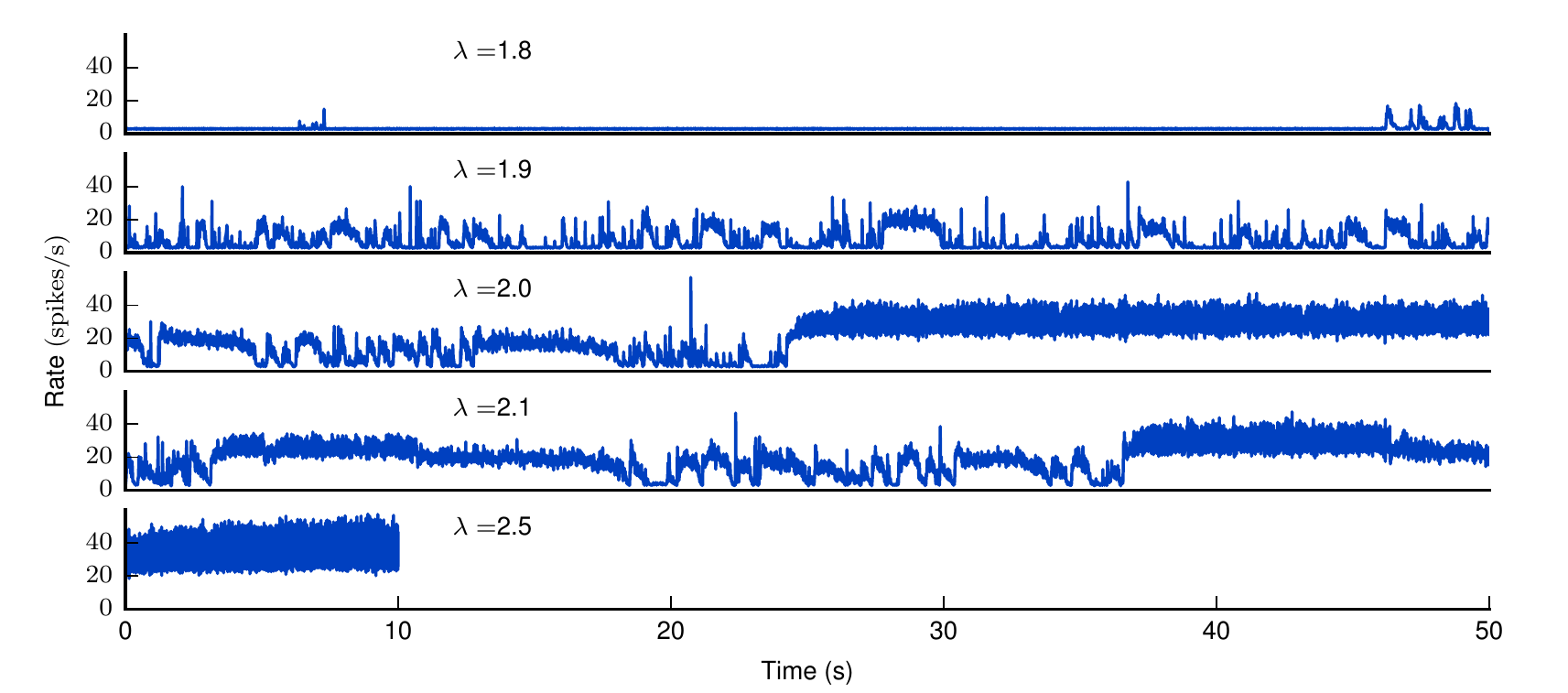}\caption{Related to \prettyref{fig:interactions}.\textbf{ Increasing cortico-cortical
synaptic weights leads to switching to a high-activity state}. Area-averaged
firing rates in V1 for four different settings of $\lambda$. The
simulation for $\lambda=2.5$ was run for $10\protect\s$ biological
time only. From$\lambda=2$ on, the network spontaneously enters a
high-activity state. For $\lambda=2.5$, the network is in this state
from the outset.}
\label{supfig:LA-HA-switching}
\end{figure}
\begin{figure}[H]
\includegraphics{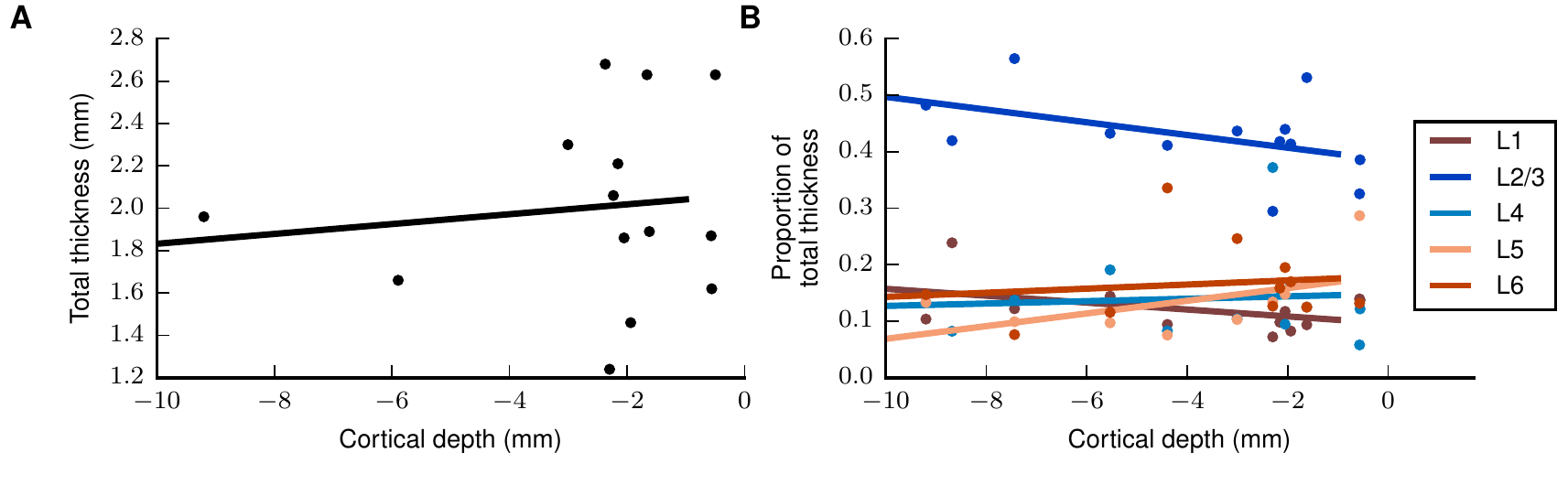}\caption{Related to \prettyref{fig:construction}. \textbf{Thickness versus
cortical depth. }(A) Total thickness vs. cortical depth and linear
least-squares fit showing no significant correlation ($r=0.12,\:p=0.68$).
(B) Relative laminar thicknesses vs. cortical depth and linear least-squares
fits also showing no significant correlation (L1: $r=-0.43,\,p=0.14$,
L2/3: $r=-0.46,\,p=0.11$, L4: $r=0.08,\:p=0.79$; L5: $r=-0.53,\:p=0.09$,
L6: $r=0.14,\,p=0.69$). The thickness data is the same as in \prettyref{fig:construction}.
Cortical depth data obtained from F99 surface statistics available
through the Caret Software \citep{vanEssen12}. Values for each area
are averaged across cortical surface and both hemispheres. The data
is obtained using the F99 Sulcal depth tool on http://cocomac.g-node.org
and can be directly accessed via these two links: \protect\url{http://cocomac.g-node.org/cocomac2/services/f99_sulcal_depth.php?atlas=FV91&shape=Depth-Right&mode=avg&output=tsv&run=1}
and \protect\url{http://cocomac.g-node.org/cocomac2/services/f99_sulcal_depth.php?atlas=FV91&shape=Depth-Left&mode=avg&output=tsv&run=1}.}
\label{supfig:thickness_vs_depth}

\end{figure}
\newpage{}
\par\end{center}

\section{Supplemental Experimental Procedures\label{sec:Supplement-Exp}}

\subsection{Cortical areas in the model}

\begin{table}[H]
\begin{centering}
\begin{tabular}{lcl}
\textbf{Lobe} & \textbf{Abbreviation} & \textbf{Brain Region}\tabularnewline
\hline 
Occipital & V1 & Visual area 1\tabularnewline
 & V2 & Visual area 2\tabularnewline
 & V3 & Visual area 3\tabularnewline
 & VP & Ventral posterior\tabularnewline
 & V3A & Visual area V3A\tabularnewline
 & V4 & Visual area 4\tabularnewline
 & VOT & Ventral occipitotemporal\tabularnewline
 & V4t & V4 transitional\tabularnewline
 & MT & Middle temporal\tabularnewline
Temporal & FST & Floor of superior temporal\tabularnewline
 & PITd & Posterior inferotemporal (dorsal)\tabularnewline
 & PITv & Posterior inferotemporal (ventral)\tabularnewline
 & CITd & Central inferotemporal (dorsal)\tabularnewline
 & CITv & Central inferotemporal (ventral)\tabularnewline
 & AITd & Anterior inferotemporal (dorsal)\tabularnewline
 & AITv & Anterior inferotemporal (ventral)\tabularnewline
 & STPp & Superior temporal polysensory (posterior)\tabularnewline
 & STPa & Superior temporal polysensory (anterior)\tabularnewline
 & TF & Parahippocampal area TF\tabularnewline
 & TH & Parahippocampal area TH\tabularnewline
Parietal & MSTd & Medial superior temporal (dorsal)\tabularnewline
 & MSTl & Medial superior temporal (lateral)\tabularnewline
 & PO & Parieto-occipital\tabularnewline
 & PIP & Posterior intraparietal\tabularnewline
 & LIP & Lateral intraparietal\tabularnewline
 & VIP & Ventral intraparietal\tabularnewline
 & MIP & Medial intraparietal\tabularnewline
 & MDP & Medial dorsal parietal\tabularnewline
 & DP & Dorsal prelunate\tabularnewline
 & 7a & 7a\tabularnewline
Frontal & FEF & Frontal eye field\tabularnewline
 & 46 & Middle frontal area 46\tabularnewline
\end{tabular}
\par\end{centering}

\caption{List of areas in the model. All vision-related areas of macaque cortex
in the parcellation of \citet{Felleman91_1}.}
\label{suptab:area_list}
\end{table}

\subsection{Inter-areal distances}

\begin{table}[H]

\input{table_distances}\caption{Distances (in mm) between the areas of the model computed as the median
of the distances between all vertex pairs of the two areas in their
surface representation in F99 space, a standard macaque cortical surface
included with Caret \citep{VanEssen01_443}, where the vertex-to-vertex
distance is the length of the shortest possible path without crossing
the cortical surface \citep{Bojak11}.}

\end{table}

\subsection{Neuron and synapse parameters}

\begin{table}[H]
\begin{centering}
\begin{tabular}{@{\hspace*{1mm}}p{1.4cm}@{}||@{\hspace*{1mm}}p{1.4cm}@{}@{\hspace*{1mm}}p{1.1cm}@{}||@{\hspace*{1mm}}p{1.1cm}@{}||@{\hspace*{1mm}}p{1.1cm}@{}@{\hspace*{1mm}}p{1.1cm}||@{\hspace*{1mm}}p{1.1cm}||@{\hspace*{1mm}}p{1.1cm}||@{\hspace*{1mm}}p{1.1cm}||@{\hspace*{1mm}}p{1.1cm}||@{\hspace*{1mm}}p{1.1cm}}
\hline 
\multicolumn{11}{>{\columncolor{parametergray}}c}{\textbf{Synapse parameters}}\tabularnewline
\multicolumn{2}{@{\hspace*{1mm}}p{2.7cm}@{}}{\textbf{Name}} & \multicolumn{3}{@{\hspace*{1mm}}p{3.65cm}@{}}{\textbf{Value}} & \multicolumn{6}{@{\hspace*{1mm}}p{6.6cm}}{\textbf{Description}}\tabularnewline
\hline 
\multicolumn{2}{>{\raggedright}p{1.35cm}}{$J\pm\delta J$} & \multicolumn{3}{>{\raggedright}p{4.3cm}}{Intra-areal connections: $87.8\pm8.8\pA$,\newline cortico-cortical
connections scaled as $J_{\mathrm{cc}}=\lambda J$, $\lambda\in[1,2.1]$\newline
cortico-cortical connections onto inhibitory populations in addition
scaled as $J_{\mathrm{cc}}^{\mathcal{I}}=\lambda_{\mathcal{I}}J_{\mathrm{cc}}^{\mathrm{\mathcal{E}}}$
, $\lambda\in\{1,2\}$} & \multicolumn{6}{>{\raggedright}p{5.1cm}}{excitatory synaptic strength}\tabularnewline
\multicolumn{2}{l}{$g$} & \multicolumn{3}{@{\hspace*{1mm}}p{3.3cm}@{}}{variable, \mbox{$g\in[-12,-4]$} } & \multicolumn{6}{@{\hspace*{1mm}}p{6.6cm}@{}}{relative inhibitory synaptic strength}\tabularnewline
\multicolumn{2}{l}{$d_{\mathrm{e}}\pm\delta d_{\mathrm{e}}$} & \multicolumn{3}{@{\hspace*{1mm}}p{3.3cm}@{}}{$1.5\pm0.75\ms$} & \multicolumn{6}{@{\hspace*{1mm}}p{6.6cm}@{}}{local excitatory transmission delay}\tabularnewline
\multicolumn{2}{l}{$d_{\mathrm{i}}\pm\delta d_{\mathrm{i}}$} & \multicolumn{3}{@{\hspace*{1mm}}p{3.3cm}@{}}{$0.75\pm0.375\ms$} & \multicolumn{6}{@{\hspace*{1mm}}p{6.6cm}@{}}{local inhibitory transmission delay}\tabularnewline
\multicolumn{2}{l}{$d\pm\delta d$} & \multicolumn{3}{@{\hspace*{1mm}}p{3.3cm}@{}}{$d=s/v_{\mathrm{t}}\pm\frac{1}{2}s/v_{\mathrm{t}}$} & \multicolumn{6}{@{\hspace*{1mm}}p{6.6cm}@{}}{inter-areal transmission delay, with $s$ the distance between areas}\tabularnewline
\multicolumn{2}{l}{$v_{\mathrm{t}}$} & \multicolumn{3}{@{\hspace*{1mm}}p{3.3cm}@{}}{$3.5\,\mathrm{m/s}$} & \multicolumn{6}{@{\hspace*{1mm}}p{6.6cm}@{}}{transmission speed}\tabularnewline
\end{tabular}
\par\end{centering}

\begin{centering}
\begin{tabular}{@{\hspace*{1mm}}p{2.7cm}@{}@{\hspace*{1mm}}p{3.65cm}@{}@{\hspace*{1mm}}p{6.6cm}}
\hline 
\multicolumn{3}{>{\columncolor{parametergray}}c}{\textbf{Neuron model}}\tabularnewline
\textbf{Name} & \textbf{Value } & \textbf{Description}\tabularnewline
\hline 
$\taum$ & $10\ms$ & membrane time constant\tabularnewline
$\taur$ & $2\ms$ & absolute refractory period\tabularnewline
$\taus$ & $0.5\ms$  & postsynaptic current time constant\tabularnewline
$C_{\mathrm{m}}$ & $250\pF$  & membrane capacity\tabularnewline
$\Vr$ & $-65\mV$ & reset potential\tabularnewline
$\theta$ & $-50\mV$ & fixed firing threshold\tabularnewline
$\EL$ & $-65\mV$ & leak potential\tabularnewline
\end{tabular}
\par\end{centering}

\caption{Parameter specification for single synapses and neurons.}
\label{suptab:Simulation-parameters}
\end{table}

\subsection{Translation of Table 4 of \citet{Hilgetag15_submitted}}

\begin{table}[H]
\begin{centering}
\begin{tabular}{cc|cc}
\textbf{Area in }\citet{Hilgetag15_submitted} & \textbf{FV91 area} & \textbf{Area in }\citet{Hilgetag15_submitted} & \textbf{FV91 area}\tabularnewline
\hline 
V1 & V1 & MST & MSTd, MSTl\tabularnewline
V2 & V2 & PIP & PIP\tabularnewline
V3 & V3 & PIT & PITd, PITv\tabularnewline
VP & VP & PO & PO\tabularnewline
MT & MT & TF & TF\tabularnewline
V3A & V3A & VIP & VIP\tabularnewline
V4 & V4 & A46v & 46\tabularnewline
V4t & V4t & A7a & 7a\tabularnewline
VOT & VOT & AIT & AITd, AITv\tabularnewline
CIT & CITd, CITv & FST & FST\tabularnewline
DP & DP & STP & STPa, STPp\tabularnewline
FEF & FEF & TH & TH\tabularnewline
LIPd, LIPv & LIP & TEO$^{\ast}$ & PITd, PITv, VOT\tabularnewline
TEr$^{\ast}$ & AITd, AITv, CITd, CITv &  & \tabularnewline
\end{tabular}
\par\end{centering}

\caption{Scheme for translating architectural types, overall neuron densities
and cortical thicknesses given in Table 4 of \citet{Hilgetag15_submitted}
to the modeled areas in the parcellation scheme of \citet{Felleman91_1}.
Entries marked with a star are used to translate the overall neuron
density and cortical thickness which are not available in the finer
of the two parcellations used by \citet{Hilgetag15_submitted}.}

\label{suptab:translate_arch_types}
\end{table}

\subsection{Relative laminar thicknesses from experimental literature}

\begin{table}[H]
\begin{centering}
\input{table_raw_laminar_thicknesses}
\par\end{centering}

\caption{Relative laminar thicknesses determined from the anatomical studies
given in the last column.}
\label{suptab:raw_laminar_thicknesses}
\end{table}

\subsection{Laminar thicknesses}

\begin{table}[H]
\begin{centering}
\input{table_laminar_thicknesses}
\par\end{centering}

\caption{Laminar thicknesses in mm for all 32 areas of the model. Values are
rounded to two decimal places. These values are used to determine
population sizes for the modeled layers 2/3, 4, 5 and 6 and to distribute
synapses across layers 1 to 6 of target areas for cortico-cortical
connections (cf. \nameref{sec:results} and \prettyref{suptab:synapse_to_cell_body}).}
\label{suptab:laminar_thicknesses}
\end{table}

\subsection{Area surfaces}

\begin{table}[H]
\begin{centering}
\begin{tabular}{cc|cc|cc}
\textbf{Area} & \textbf{Surface area ($\mathrm{mm}^{2}$)} & \textbf{Area} & \textbf{Surface area ($\mathrm{mm}^{2}$)} & \textbf{Area} & \textbf{Surface area ($\mathrm{mm}^{2}$)}\tabularnewline
\hline 
V1 & 1484.63 & V3 & 120.57 & PO & 75.37\tabularnewline
V2 & 1193.40 & CITv & 114.67 & VOT & 70.11\tabularnewline
V4 & 561.41 & DP & 113.83 & LIP & 56.04\tabularnewline
STPp & 245.48 & PIP & 106.15 & MT & 55.90\tabularnewline
TF & 197.40 & PITv & 100.34 & FST & 61.33\tabularnewline
46 & 185.16 & AITd & 91.59 & CITd & 57.54\tabularnewline
FEF & 161.54 & VIP & 85.06 & MIP & 45.09\tabularnewline
7a & 157.34 & V3A & 96.96 & TH & 44.60\tabularnewline
PITd & 145.38 & AITv & 93.12 & MSTl & 29.19\tabularnewline
VP & 130.58 & STPa & 78.72 & V4t & 28.23\tabularnewline
MSTd & 120.57 & MDP & 77.49 &  & \tabularnewline
\end{tabular}
\par\end{centering}

\caption{Surface areas computed with Caret \citep{VanEssen01_443} on the basis
of each area's representation on the F99 cortical surface \citep{VanEssen02_574}.
Areas are ordered from large to small.}

\label{suptab:Area_surfaces}
\end{table}

\subsection{Population sizes}

\begin{table}[H]
\begin{centering}
\input{table_neuron_numbers}
\par\end{centering}

\caption{Estimated population sizes across layers and areas underneath $1\,\mathrm{mm^{2}}$
of cortical surface in each area.}
 \label{suptab:neuron_numbers}
\end{table}

\subsection*{Derivation of the conversion factor $c_{A}(R)$ for the local connectivity}

The indegrees of the microcircuit model \citep{Potjans14_785} $K_{ij}^{\mathrm{\prime}}(R)$
are adapted to the area-specific laminar compositions of the multi-area
model with an area-specific factor $c_{A}(R)$,
\[
K_{ij}(R)=c_{A}(R)K_{ij}^{\mathrm{\prime}}(R)\,\,\,\forall i,j\,,
\]
where $i,j$ denote single populations in the $1\mm^{2}$ patch of
the cortical area. The total number of synapses local to the patch
(type I) is the sum over the projections between all populations of
the area:

\[
N_{\mathrm{syn,I}}=\sum_{i,j}N_{i}K_{ij}=c_{A}\sum_{i,j}N_{i}K_{ij}^{\prime}\,.
\]
We thus obtain $c_{A}(R)$ by determining $N_{\mathrm{syn,I}}$. To
this end, we use retrograde tracing data from \citet{Markov11_1254}
consisting of fractions of labeled neurons ($FLN$) per area as a
result of injections into one area at a time. The fraction intrinsic
to the injected area, $FLN_{\mathrm{i}}$, is approximately equal
for all 9 areas where this fraction was determined, with a mean of
$0.79$. For areas modeled with reduced size, this fraction is smaller
because, in that case, synapses of both type I and II contribute to
the value of $0.79$ (\prettyref{fig:construction}E). We approximate
the increasing contribution of type I synapses with the modeled area
size as the increase in indegrees averaged over population pairs,
\begin{eqnarray*}
\frac{N_{\mathrm{syn,I}}(R)/N_{\mathrm{syn,tot}}(R)}{N_{\mathrm{syn,I}}(R_{\mathrm{full}})/N_{\mathrm{syn,tot}}(R_{\mathrm{full}})} & = & \Bigg\langle\frac{K_{ij}(R)}{K_{ij}(R_{\mathrm{full}})}\Bigg\rangle_{ij}=\Bigg\langle\frac{K_{ij}^{\prime}(R)}{K_{ij}^{\prime}(R_{\mathrm{full}})}\Bigg\rangle_{ij}\,,
\end{eqnarray*}
where in the last step we use \prettyref{eq:indegree_equation}. Using
$N_{\mathrm{syn,I}}(R_{\mathrm{full}})/N_{\mathrm{syn,tot}}(R_{\mathrm{full}})=FLN_{\mathrm{i}}$,
we obtain 
\[
N_{\mathrm{syn,I}}(R)=N_{\mathrm{syn,tot}}(R)\,FLN_{\mathrm{i}}\Bigg\langle\frac{K_{ij}^{\prime}(R)}{K_{ij}^{\prime}(R_{\mathrm{full}})}\Bigg\rangle_{ij}\,,
\]
where $N_{\mathrm{syn,tot}}(R)=\rho_{\mathrm{syn}}\pi R^{2}D$ with
$D$ the total thickness of the given area. The conversion factor
can thus be obtained with
\[
c_{A}(R)=\frac{N_{\mathrm{syn,tot}}(R)}{\sum_{i,j}N_{i}K_{ij}^{\prime}}\,FLN_{\mathrm{i}}\Bigg\langle\frac{K_{ij}^{\prime}(R)}{K_{ij}^{\prime}(R_{\mathrm{full}})}\Bigg\rangle_{ij}\,.
\]
 We substitute this into \prettyref{eq:indegree_equation} for the
modeled areas where $R=R_{0}$ and obtain the population-specific
indegrees for type I synapses:

\[
K_{ij,\mathrm{I}}\defeq K_{ij}\left(R=R_{0}\right)
\]

\subsection{Processing of CoCoMac data}

We use a new release of CoCoMac, in which mappings from brain regions
in other nomenclatures were scrutinized to ensure a consistent transfer
of connections into the FV91 name space. The CoCoMac database provides
information on laminar patterns on the source side from retrograde
tracing studies as well as on the target side from anterograde tracing
studies. The data was extracted by using the following link, which
specifies all search options: \url{http://cocomac.g-node.org/cocomac2/services/connectivity_matrix.php?dbdate=20141022&AP=AxonalProjections_FV91&constraint=&origins=&terminals=&square=1&merge=max&laminar=both&format=json&cite=1}

Furthermore, we obtained the numbers of confirmative studies for each
area-level connection with the following link: \url{http://cocomac.g-node.org/cocomac2/services/connectivity_matrix.php?dbdate=20141022&AP=AxonalProjections_FV91&constraint=&origins=&terminals=&square=1&merge=count&laminar=off&format=json&cite=1}

To process these data, we applied the following steps:
\begin{itemize}
\item A connection is assumed to exist if there is at least one confirmative
study reporting it.
\item A connection from layer 2/3 is modeled if CoCoMac indicates a connection
from either or both of layers 2 and 3.
\item In the database, some layers carry an `X' indicating a connection
of unknown strength. We interpret these as `2' (corresponding to medium
connection strength).
\item We take connection strengths in CoCoMac to represent numbers of synapses
in orders of magnitude, i.e., the relative number of synapses $N_{\mathrm{syn}}^{\nu}$
in layer $\nu$ of area $A$ with connection strength $s(\nu)$ is
computed as $N_{\mathrm{syn}}^{\nu}=10^{s(v)}/\sum_{v^{\prime}\in A}10^{s(\nu^{\prime})}$.
\end{itemize}

\subsection{Mapping of injection sites to FV91 parcellation}

\begin{table}[H]
\begin{centering}
\begin{tabular}{ccc|ccc}
\textbf{Monkey} & \textbf{M132 area} & \textbf{FV91 area} & \textbf{Monkey} & \textbf{M132 area} & \textbf{FV91 area}\tabularnewline
\hline 
M88RH & V1 & V1 & M101LH & V2 & V2\tabularnewline
M121LH & V1 & V1 & M101RH & V2 & V2\tabularnewline
M81LH & V1 & V1 & M103LH & V2 & V2\tabularnewline
M85LH & V1 & V1 & M123LH & V4 & V4\tabularnewline
M85RH & V1 & V1 & M121RH & V4 & V4\tabularnewline
BB289RH & STPr & STPa & M119LH & TEO & V4\tabularnewline
BB289LH & STPi & STPp & BB135LH & 7A & 7a\tabularnewline
M90RH & STPc & STPp & M89LH & DP & DP\tabularnewline
M106LH & 9/46d & FEF & BB272RH & 8l & FEF\tabularnewline
M133LH & MT & MSTd & M116LH & 46d & 46\tabularnewline
M116RH & 9/46v & 46 & BB272LH & 8m & FEF\tabularnewline
M128RH & TEPd & CITv & M108LH & PBr & STPp\tabularnewline
\end{tabular}
\par\end{centering}

\caption{Injected areas of the data set of \citet{Markov2014_17} in the M132
parcellation and corresponding areas in the FV91 scheme. Only the
injections in vision-related cortex are shown.}
\label{suptab:Injected-areas-markov}
\end{table}

\subsection{Mapping of synapse to cell-body locations}

Detailed calculation in section \nameref{sec:exp_procedures}. The
numbers are listed in \prettyref{suptab:synapse_to_cell_body}.

\begin{table}[H]
\centering{}%
\begin{tabular}{cc|ccccc}
 &  & \multicolumn{5}{c}{\textbf{Synapse layer}}\tabularnewline
\hline 
\multirow{9}{*}{\begin{turn}{90}
\textbf{Target population}
\end{turn}} &  & \textbf{1} & \textbf{2/3} & \textbf{4} & \textbf{5} & \textbf{6}\tabularnewline
 & 2/3E & 0.57 &  &  &  & \tabularnewline
 & 2/3I &  & 0.16 &  &  & \tabularnewline
 & 4E & 0.18 & 0.84 & 0.73 &  & \tabularnewline
 & 4I &  &  & 0.16 &  & \tabularnewline
 & 5E & 0.25 &  & 0.02 & 0.76 & \tabularnewline
 & 5I &  &  &  & 0.1 & \tabularnewline
 & 6E & 0.003 &  & 0.09 & 0.14 & 0.85\tabularnewline
 & 6I &  &  &  &  & 0.15\tabularnewline
\end{tabular}\caption{Conditional probabilities $\mathcal{P}(i|s_{\mathrm{cc}}\in v)$ for
the target neuron to belong to population $i$ if a cortico-cortical
synapse $s_{\mathrm{cc}}$ is located in layer $v$, computed with
\prettyref{eq:synapse_to_cell_body} applied to the data set of \citet{Binzegger04}.
Empty cells signal zero probabilities. }
\label{suptab:synapse_to_cell_body}
\end{table}

\subsection{External input}

\begin{table}[H]
\begin{centering}
\input{table_external_indegrees}
\par\end{centering}

\caption{Numbers of extrinsic synapses per neuron for all areas of the model
with $\kappa=1.125$.}
\label{suptab:external-input}
\end{table}

\subsection{Analytical mean-field theory}

In \citet{Schuecker15_arxiv}, analytical mean-field theory is derived
describing the stationary population-averaged firing rates of the
model. In the diffusion approximation, which is valid for high indegrees
and small synaptic weights, the dynamics of the membrane potential
$V$ and synaptic current $I_{\mathrm{s}}$are described by \citep{Fourcaud02}
\begin{eqnarray*}
\taum\frac{\d V}{\d t} & = & -V+I_{\mathrm{s}}(t)\\
\taus\frac{\d I_{\mathrm{s}}}{\d t} & = & -I_{\mathrm{s}}+\mu+\sigma\sqrt{\taum}\xi(t),
\end{eqnarray*}
where the input spike trains are replaced by a current fluctuating
around the mean $\mu$ with variance $\sigma$ with fluctuations drawn
from a random Gaussian process $\xi(t)$ with $\langle\xi(t)\rangle=0$
and $\langle\xi(t)\xi(t^{\prime})\rangle=\delta(t-t^{\prime})$. Going
from the single-neuron level to a description of populations, we define
the population-averaged firing rate $\nu_{i}$ due to the population-specific
input $\mu_{i},\,\sigma_{i}$. The stationary firing rates $\nu_{i}$
are then given by \citep{Fourcaud02}

\begin{eqnarray*}
\frac{1}{\nu_{i}} & = & \taur+\taum\sqrt{\pi}\int_{\frac{V_{r}-\mu_{i}(\boldsymbol{A})}{\sigma_{i}(\boldsymbol{A})}+\gamma\sqrt{\frac{\taus}{\taum}}}^{\frac{\Theta-\mu_{i}(\boldsymbol{A})}{\sigma_{i}(\boldsymbol{A})}+\gamma\sqrt{\frac{\taus}{\taum}}}e^{x^{2}}\ (1+\erf(x))\,\d x\\
 & \eqqcolon & 1/\Phi_{i}(\boldsymbol{\nu},\boldsymbol{A})\\
\mu_{i}(\boldsymbol{A}) & = & \taum\sum_{j}K_{ij}J{}_{ij}\nu_{j}+\taum K_{\mathrm{ext}}J_{\mathrm{ext}}\nuext\\
\sigma_{i}^{2}(\boldsymbol{A}) & = & \taum\sum_{j}K_{ij}J_{ij}^{2}\nu_{j}+\taum K_{\mathrm{ext}}J_{\mathrm{ext}}^{2}\nuext,
\end{eqnarray*}
which holds up to linear order in $\sqrt{\taus/\taum}$ and where
$\gamma=|\zeta(1/2)|/\sqrt{2},$ with $\zeta$ denoting the Riemann
zeta function \citep{Abramowitz74}.

\subsection{Algorithm for the temporal hierarchy}

To determine a temporal hierarchy for the onset of population bursts,
we determine the peak locations $\tau_{AB}$ of the cross-correlation
function for each pair of areas $A,\,B$. We then define a scalar
function for the deviation between the distance of hierarchical levels
$h(A),\,h(B)$ and peak locations,
\[
f(A,B)=h(A)-h(B)-\tau_{AB}\,.
\]
To determine the hierarchical levels, we minimize the sum of $f(A,B)$
over all pairs of areas,
\[
S=\sum_{A,B}f(A,B)\,,
\]
using the \texttt{optimize.minimize} function of the \texttt{scipy}
library \citep{scipy01} with random initial hierarchical levels.
We verified that the initial choice of hierarchical levels does not
influence the final result. We obtain hierarchical levels on an arbitrary
scale, which we normalize to values $h(A)\in[0,1]\,\forall A$.

%% file: table_distances.tex
\footnotesize\begin{center} 
\begin{tabular}{l|llllllllllllllll} 
\textbf{Area} & \textbf{V1} & \textbf{V2} & \textbf{VP} & \textbf{V3} & \textbf{V3A} & \textbf{MT} & \textbf{V4t} & \textbf{V4} & \textbf{VOT} & \textbf{MSTd} & \textbf{PIP} & \textbf{PO} & \textbf{DP} & \textbf{MIP} & \textbf{MDP} & \textbf{VIP} \\ \hline 
\textbf{V1} & 0.0 & 17.9 & 19.9 & 14.6 & 16.8 & 22.5 & 23.1 & 22.9 & 29.0 & 26.8 & 18.8 & 21.5 & 23.7 & 24.5 & 29.2 & 26.3 \\ 
\textbf{V2} & 17.9 & 0.0 & 16.1 & 17.8 & 18.2 & 20.0 & 20.5 & 21.2 & 24.5 & 24.4 & 19.8 & 23.8 & 24.6 & 26.0 & 30.8 & 25.8 \\ 
\textbf{VP} & 19.9 & 16.1 & 0.0 & 20.8 & 19.0 & 14.9 & 15.1 & 14.6 & 12.8 & 20.9 & 20.1 & 25.2 & 25.4 & 26.9 & 31.9 & 25.5 \\ 
\textbf{V3} & 14.6 & 17.8 & 20.8 & 0.0 & 8.1 & 15.9 & 17.0 & 18.5 & 26.9 & 19.4 & 10.6 & 14.6 & 15.1 & 16.9 & 22.0 & 18.1 \\ 
\textbf{V3A} & 16.8 & 18.2 & 19.0 & 8.1 & 0.0 & 12.4 & 13.4 & 15.6 & 23.4 & 15.4 & 9.2 & 15.0 & 9.4 & 16.3 & 21.2 & 13.8 \\ 
\textbf{MT} & 22.5 & 20.0 & 14.9 & 15.9 & 12.4 & 0.0 & 6.0 & 11.4 & 13.7 & 10.0 & 13.2 & 19.1 & 16.6 & 20.0 & 23.9 & 13.7 \\ 
\textbf{V4t} & 23.1 & 20.5 & 15.1 & 17.0 & 13.4 & 6.0 & 0.0 & 9.9 & 12.1 & 12.1 & 15.2 & 21.3 & 17.8 & 22.1 & 26.3 & 16.4 \\ 
\textbf{V4} & 22.9 & 21.2 & 14.6 & 18.5 & 15.6 & 11.4 & 9.9 & 0.0 & 13.1 & 17.8 & 18.6 & 24.6 & 20.4 & 25.6 & 30.5 & 21.4 \\ 
\textbf{VOT} & 29.0 & 24.5 & 12.8 & 26.9 & 23.4 & 13.7 & 12.1 & 13.1 & 0.0 & 19.7 & 24.6 & 30.0 & 28.5 & 31.0 & 36.0 & 26.6 \\ 
\textbf{MSTd} & 26.8 & 24.4 & 20.9 & 19.4 & 15.4 & 10.0 & 12.1 & 17.8 & 19.7 & 0.0 & 14.5 & 20.6 & 17.1 & 20.2 & 24.1 & 11.5 \\ 
\textbf{PIP} & 18.8 & 19.8 & 20.1 & 10.6 & 9.2 & 13.2 & 15.2 & 18.6 & 24.6 & 14.5 & 0.0 & 9.5 & 12.3 & 9.8 & 14.2 & 11.2 \\ 
\textbf{PO} & 21.5 & 23.8 & 25.2 & 14.6 & 15.0 & 19.1 & 21.3 & 24.6 & 30.0 & 20.6 & 9.5 & 0.0 & 18.9 & 6.9 & 10.0 & 16.0 \\ 
\textbf{DP} & 23.7 & 24.6 & 25.4 & 15.1 & 9.4 & 16.6 & 17.8 & 20.4 & 28.5 & 17.1 & 12.3 & 18.9 & 0.0 & 18.3 & 22.0 & 11.1 \\ 
\textbf{MIP} & 24.5 & 26.0 & 26.9 & 16.9 & 16.3 & 20.0 & 22.1 & 25.6 & 31.0 & 20.2 & 9.8 & 6.9 & 18.3 & 0.0 & 6.3 & 14.6 \\ 
\textbf{MDP} & 29.2 & 30.8 & 31.9 & 22.0 & 21.2 & 23.9 & 26.3 & 30.5 & 36.0 & 24.1 & 14.2 & 10.0 & 22.0 & 6.3 & 0.0 & 17.4 \\ 
\textbf{VIP} & 26.3 & 25.8 & 25.5 & 18.1 & 13.8 & 13.7 & 16.4 & 21.4 & 26.6 & 11.5 & 11.2 & 16.0 & 11.1 & 14.6 & 17.4 & 0.0 \\ 
\textbf{LIP} & 27.8 & 27.6 & 28.0 & 19.3 & 14.6 & 16.0 & 18.6 & 23.3 & 29.1 & 12.4 & 13.6 & 19.5 & 10.1 & 18.2 & 21.1 & 7.4 \\ 
\textbf{PITv} & 32.9 & 28.0 & 16.8 & 30.4 & 26.8 & 16.3 & 14.8 & 16.3 & 7.4 & 21.5 & 27.6 & 32.9 & 31.9 & 34.0 & 38.7 & 28.8 \\ 
\textbf{PITd} & 31.6 & 27.3 & 17.8 & 27.4 & 23.5 & 13.1 & 11.6 & 14.1 & 8.6 & 18.7 & 24.8 & 30.6 & 28.2 & 31.4 & 35.9 & 25.7 \\ 
\textbf{MSTl} & 28.4 & 24.4 & 17.4 & 22.6 & 18.9 & 8.2 & 9.9 & 15.7 & 13.2 & 9.1 & 18.0 & 24.0 & 22.7 & 24.2 & 28.3 & 16.9 \\ 
\textbf{CITv} & 38.8 & 33.4 & 22.4 & 36.0 & 32.7 & 21.6 & 21.1 & 22.5 & 12.9 & 25.7 & 32.4 & 37.6 & 37.7 & 38.8 & 43.1 & 33.1 \\ 
\textbf{CITd} & 37.7 & 32.5 & 22.0 & 34.3 & 31.0 & 19.5 & 19.2 & 20.9 & 12.2 & 23.2 & 30.5 & 35.9 & 35.7 & 37.0 & 41.0 & 30.7 \\ 
\textbf{FEF} & 57.1 & 53.9 & 48.3 & 50.0 & 45.8 & 37.9 & 40.3 & 45.8 & 42.1 & 36.4 & 42.9 & 47.9 & 46.2 & 45.4 & 46.7 & 36.8 \\ 
\textbf{TF} & 29.6 & 24.8 & 16.3 & 27.4 & 24.7 & 15.8 & 17.1 & 20.1 & 14.4 & 20.2 & 23.5 & 28.0 & 29.8 & 29.5 & 33.7 & 25.9 \\ 
\textbf{AITv} & 43.8 & 38.2 & 28.2 & 41.1 & 38.1 & 26.6 & 27.0 & 28.9 & 19.5 & 30.1 & 36.7 & 41.4 & 42.7 & 42.5 & 46.7 & 36.5 \\ 
\textbf{FST} & 33.7 & 28.9 & 20.3 & 28.8 & 25.6 & 14.2 & 15.7 & 19.2 & 12.8 & 16.5 & 24.6 & 29.8 & 29.6 & 30.6 & 34.2 & 23.8 \\ 
\textbf{7a} & 28.2 & 27.6 & 27.4 & 19.9 & 14.5 & 15.5 & 18.0 & 22.4 & 27.8 & 11.0 & 14.5 & 20.9 & 11.5 & 20.0 & 23.2 & 9.4 \\ 
\textbf{STPp} & 38.0 & 34.3 & 27.7 & 31.7 & 28.1 & 18.1 & 20.0 & 25.5 & 22.5 & 16.0 & 27.2 & 32.9 & 30.8 & 33.0 & 36.6 & 24.3 \\ 
\textbf{STPa} & 44.3 & 39.2 & 30.2 & 40.2 & 37.1 & 25.5 & 27.0 & 30.0 & 21.8 & 27.3 & 35.3 & 40.4 & 40.9 & 41.0 & 44.5 & 33.6 \\ 
\textbf{46} & 62.9 & 59.5 & 54.1 & 55.9 & 51.8 & 43.9 & 46.4 & 51.5 & 47.7 & 42.4 & 49.0 & 53.3 & 52.0 & 50.8 & 52.0 & 42.6 \\ 
\textbf{AITd} & 46.3 & 40.9 & 31.2 & 43.3 & 40.4 & 28.4 & 29.0 & 30.3 & 21.6 & 31.6 & 39.0 & 43.9 & 44.7 & 44.7 & 48.9 & 38.1 \\ 
\textbf{TH} & 30.8 & 26.3 & 19.9 & 27.5 & 25.1 & 17.1 & 18.9 & 22.6 & 18.1 & 20.6 & 22.4 & 26.2 & 29.4 & 28.2 & 31.4 & 24.9 \\ 
&&&&&&&&&&&&&&&& \\ 
\textbf{Area} & \textbf{LIP} & \textbf{PITv} & \textbf{PITd} & \textbf{MSTl} & \textbf{CITv} & \textbf{CITd} & \textbf{FEF} & \textbf{TF} & \textbf{AITv} & \textbf{FST} & \textbf{7a} & \textbf{STPp} & \textbf{STPa} & \textbf{46} & \textbf{AITd} & \textbf{TH} \\ \hline 
\textbf{V1} & 27.8 & 32.9 & 31.6 & 28.4 & 38.8 & 37.7 & 57.1 & 29.6 & 43.8 & 33.7 & 28.2 & 38.0 & 44.3 & 62.9 & 46.3 & 30.8 \\ 
\textbf{V2} & 27.6 & 28.0 & 27.3 & 24.4 & 33.4 & 32.5 & 53.9 & 24.8 & 38.2 & 28.9 & 27.6 & 34.3 & 39.2 & 59.5 & 40.9 & 26.3 \\ 
\textbf{VP} & 28.0 & 16.8 & 17.8 & 17.4 & 22.4 & 22.0 & 48.3 & 16.3 & 28.2 & 20.3 & 27.4 & 27.7 & 30.2 & 54.1 & 31.2 & 19.9 \\ 
\textbf{V3} & 19.3 & 30.4 & 27.4 & 22.6 & 36.0 & 34.3 & 50.0 & 27.4 & 41.1 & 28.8 & 19.9 & 31.7 & 40.2 & 55.9 & 43.3 & 27.5 \\ 
\textbf{V3A} & 14.6 & 26.8 & 23.5 & 18.9 & 32.7 & 31.0 & 45.8 & 24.7 & 38.1 & 25.6 & 14.5 & 28.1 & 37.1 & 51.8 & 40.4 & 25.1 \\ 
\textbf{MT} & 16.0 & 16.3 & 13.1 & 8.2 & 21.6 & 19.5 & 37.9 & 15.8 & 26.6 & 14.2 & 15.5 & 18.1 & 25.5 & 43.9 & 28.4 & 17.1 \\ 
\textbf{V4t} & 18.6 & 14.8 & 11.6 & 9.9 & 21.1 & 19.2 & 40.3 & 17.1 & 27.0 & 15.7 & 18.0 & 20.0 & 27.0 & 46.4 & 29.0 & 18.9 \\ 
\textbf{V4} & 23.3 & 16.3 & 14.1 & 15.7 & 22.5 & 20.9 & 45.8 & 20.1 & 28.9 & 19.2 & 22.4 & 25.5 & 30.0 & 51.5 & 30.3 & 22.6 \\ 
\textbf{VOT} & 29.1 & 7.4 & 8.6 & 13.2 & 12.9 & 12.2 & 42.1 & 14.4 & 19.5 & 12.8 & 27.8 & 22.5 & 21.8 & 47.7 & 21.6 & 18.1 \\ 
\textbf{MSTd} & 12.4 & 21.5 & 18.7 & 9.1 & 25.7 & 23.2 & 36.4 & 20.2 & 30.1 & 16.5 & 11.0 & 16.0 & 27.3 & 42.4 & 31.6 & 20.6 \\ 
\textbf{PIP} & 13.6 & 27.6 & 24.8 & 18.0 & 32.4 & 30.5 & 42.9 & 23.5 & 36.7 & 24.6 & 14.5 & 27.2 & 35.3 & 49.0 & 39.0 & 22.4 \\ 
\textbf{PO} & 19.5 & 32.9 & 30.6 & 24.0 & 37.6 & 35.9 & 47.9 & 28.0 & 41.4 & 29.8 & 20.9 & 32.9 & 40.4 & 53.3 & 43.9 & 26.2 \\ 
\textbf{DP} & 10.1 & 31.9 & 28.2 & 22.7 & 37.7 & 35.7 & 46.2 & 29.8 & 42.7 & 29.6 & 11.5 & 30.8 & 40.9 & 52.0 & 44.7 & 29.4 \\ 
\textbf{MIP} & 18.2 & 34.0 & 31.4 & 24.2 & 38.8 & 37.0 & 45.4 & 29.5 & 42.5 & 30.6 & 20.0 & 33.0 & 41.0 & 50.8 & 44.7 & 28.2 \\ 
\textbf{MDP} & 21.1 & 38.7 & 35.9 & 28.3 & 43.1 & 41.0 & 46.7 & 33.7 & 46.7 & 34.2 & 23.2 & 36.6 & 44.5 & 52.0 & 48.9 & 31.4 \\ 
\textbf{VIP} & 7.4 & 28.8 & 25.7 & 16.9 & 33.1 & 30.7 & 36.8 & 25.9 & 36.5 & 23.8 & 9.4 & 24.3 & 33.6 & 42.6 & 38.1 & 24.9 \\ 
\textbf{LIP} & 0.0 & 31.2 & 28.0 & 18.7 & 35.5 & 33.1 & 39.0 & 28.9 & 39.5 & 26.2 & 7.1 & 25.5 & 36.2 & 45.1 & 40.9 & 28.2 \\ 
\textbf{PITv} & 31.2 & 0.0 & 8.3 & 13.9 & 9.3 & 8.3 & 40.5 & 14.1 & 15.4 & 11.2 & 29.7 & 21.6 & 18.4 & 45.8 & 17.2 & 18.1 \\ 
\textbf{PITd} & 28.0 & 8.3 & 0.0 & 12.1 & 12.7 & 10.4 & 39.9 & 15.8 & 19.1 & 10.9 & 26.5 & 20.2 & 20.2 & 45.4 & 19.7 & 19.2 \\ 
\textbf{MSTl} & 18.7 & 13.9 & 12.1 & 0.0 & 17.7 & 15.1 & 32.4 & 14.3 & 22.2 & 8.8 & 17.2 & 11.9 & 19.8 & 38.4 & 23.9 & 15.8 \\ 
\textbf{CITv} & 35.5 & 9.3 & 12.7 & 17.7 & 0.0 & 6.4 & 38.7 & 14.6 & 9.3 & 11.8 & 33.9 & 20.9 & 13.8 & 43.2 & 10.9 & 18.3 \\ 
\textbf{CITd} & 33.1 & 8.3 & 10.4 & 15.1 & 6.4 & 0.0 & 36.5 & 13.9 & 10.3 & 9.0 & 31.4 & 18.5 & 12.4 & 41.2 & 10.7 & 17.2 \\ 
\textbf{FEF} & 39.0 & 40.5 & 39.9 & 32.4 & 38.7 & 36.5 & 0.0 & 39.9 & 36.5 & 30.5 & 39.4 & 33.3 & 29.3 & 11.2 & 35.2 & 39.8 \\ 
\textbf{TF} & 28.9 & 14.1 & 15.8 & 14.3 & 14.6 & 13.9 & 39.9 & 0.0 & 16.4 & 12.7 & 27.9 & 21.7 & 19.0 & 44.7 & 19.6 & 9.7 \\ 
\textbf{AITv} & 39.5 & 15.4 & 19.1 & 22.2 & 9.3 & 10.3 & 36.5 & 16.4 & 0.0 & 14.3 & 38.3 & 21.7 & 10.7 & 39.9 & 7.4 & 18.5 \\ 
\textbf{FST} & 26.2 & 11.2 & 10.9 & 8.8 & 11.8 & 9.0 & 30.5 & 12.7 & 14.3 & 0.0 & 24.7 & 12.4 & 12.2 & 36.0 & 15.5 & 14.6 \\ 
\textbf{7a} & 7.1 & 29.7 & 26.5 & 17.2 & 33.9 & 31.4 & 39.4 & 27.9 & 38.3 & 24.7 & 0.0 & 23.6 & 35.3 & 45.4 & 40.0 & 27.6 \\ 
\textbf{STPp} & 25.5 & 21.6 & 20.2 & 11.9 & 20.9 & 18.5 & 33.3 & 21.7 & 21.7 & 12.4 & 23.6 & 0.0 & 16.0 & 38.4 & 22.2 & 23.3 \\ 
\textbf{STPa} & 36.2 & 18.4 & 20.2 & 19.8 & 13.8 & 12.4 & 29.3 & 19.0 & 10.7 & 12.2 & 35.3 & 16.0 & 0.0 & 33.1 & 10.2 & 20.7 \\ 
\textbf{46} & 45.1 & 45.8 & 45.4 & 38.4 & 43.2 & 41.2 & 11.2 & 44.7 & 39.9 & 36.0 & 45.4 & 38.4 & 33.1 & 0.0 & 38.3 & 44.6 \\ 
\textbf{AITd} & 40.9 & 17.2 & 19.7 & 23.9 & 10.9 & 10.7 & 35.2 & 19.6 & 7.4 & 15.5 & 40.0 & 22.2 & 10.2 & 38.3 & 0.0 & 21.8 \\ 
\textbf{TH} & 28.2 & 18.1 & 19.2 & 15.8 & 18.3 & 17.2 & 39.8 & 9.7 & 18.5 & 14.6 & 27.6 & 23.3 & 20.7 & 44.6 & 21.8 & 0.0 \\ 
\end{tabular} 
\end{center} 

%% file: table_raw_laminar_thicknesses.tex
\begin{table}[H] 
\centering 
\begin{tabularx}{\textwidth}{p{1.2cm}p{1.cm}p{1.cm}p{1.cm}p{1.cm}p{1.cm}p{5cm}l} 
\textbf{Area} & \textbf{1} & \textbf{2/3} & \textbf{4} & \textbf{5} & \textbf{6} & \textbf{Source} \\ 
\hline 
V1 & 0.08 & 0.25 & 0.37 & 0.14 & 0.16 & \citet{OKusky82_278} \\ 
 V1 & 0.09 & 0.29 & 0.39 & 0.11 & 0.12 & \citet{Rakic91_2083} \\ 
 V1 & 0.08 & 0.32 & 0.38 & 0.14 & 0.08 & \citet{Felleman97_21} \\ 
 V1 & 0.05 & 0.31 & 0.36 & 0.14 & 0.14 & \citet{Eggan07_175} \\ 
 V2 & 0.07 & 0.41 & 0.14 & 0.21 & 0.18 & \citet{Markov2014_17} \\ 
 V2 & 0.1 & 0.42 & 0.19 & 0.13 & 0.16 & \citet{Rakic91_2083} \\ 
 V3 & 0.09 & 0.58 & 0.12 & 0.1 & 0.12 & \citet{Markov2014_17} \\ 
 V3 & 0.2 & 0.29 & 0.27 & nan & nan & \citet{Angelucci02} \\ 
 MT & 0.11 & 0.54 & 0.13 & 0.11 & 0.11 & \citet{Markov2014_17} \\ 
 MT & 0.09 & 0.43 & 0.14 & 0.16 & 0.18 & \citet{Preuss91_429} \\ 
 V4 & 0.09 & 0.53 & 0.12 & 0.12 & 0.12 & \citet{Rockland92_353} \\ 
 MIP & 0.09 & 0.41 & 0.08 & 0.08 & 0.34 & \citet{Rozzi06_1389} \\ 
 VIP & 0.12 & 0.56 & 0.14 & 0.1 & 0.08 & \citet{Preuss91_429} \\ 
 LIP & 0.09 & 0.36 & 0.09 & 0.08 & 0.39 & \citet{Rozzi06_1389} \\ 
 LIP & 0.13 & 0.52 & 0.12 & 0.13 & 0.1 & \citet{Preuss91_429} \\ 
 FEF & 0.1 & 0.42 & 0.16 & 0.17 & 0.16 & \citet{Boussaoud90_462} \\ 
 TF & 0.14 & 0.39 & 0.12 & nan & nan & \citet{Preuss91_429} \\ 
 FST & 0.24 & 0.42 & 0.08 & nan & nan & \citet{Lavenex02_394} \\ 
 46 & 0.1 & 0.45 & 0.1 & 0.15 & 0.2 & \citet{Eggan07_175} \\ 
 46 & 0.13 & 0.43 & 0.09 & nan & nan & \citet{Petrides99_1011} \\ 
 TH & nan & nan & 0.0 & nan & nan & \citet{Suzuki03_67} \\ 
 TH & 0.14 & 0.33 & 0.12 & 0.29 & 0.13 & \citet{Preuss91_429} \\ 
 \end{tabularx} 
\end{table} 

%% file: table_laminar_thicknesses.tex
\begin{table}[H] 
\begin{center} 
\begin{tabular}{p{1.6cm}p{1.6cm}p{1.6cm}p{1.6cm}p{1.6cm}p{1.6cm}p{1.6cm}} 
\textbf{Area} & \textbf{1} & \textbf{2/3} & \textbf{4} & \textbf{5} & \textbf{6} & \textbf{Total} \\ 
\hline 
V1 & 0.09 & 0.37 & 0.46 & 0.17 & 0.16 & 1.24 \\
V2 & 0.12 & 0.60 & 0.24 & 0.25 & 0.25 & 1.46 \\
VP & 0.18 & 0.63 & 0.32 & 0.21 & 0.25 & 1.59 \\
V3 & 0.23 & 0.70 & 0.31 & 0.16 & 0.19 & 1.59 \\
PIP & 0.26 & 0.92 & 0.24 & 0.30 & 0.36 & 2.07 \\
V3A & 0.20 & 0.71 & 0.24 & 0.23 & 0.28 & 1.66 \\
MT & 0.20 & 0.95 & 0.26 & 0.26 & 0.29 & 1.96 \\
V4t & 0.22 & 0.80 & 0.29 & 0.26 & 0.31 & 1.88 \\
V4 & 0.18 & 1.00 & 0.24 & 0.24 & 0.24 & 1.89 \\
PO & 0.26 & 0.92 & 0.24 & 0.30 & 0.36 & 2.07 \\
VOT & 0.23 & 0.81 & 0.28 & 0.27 & 0.32 & 1.90 \\
DP & 0.26 & 0.91 & 0.23 & 0.30 & 0.36 & 2.06 \\
MIP & 0.20 & 0.85 & 0.17 & 0.16 & 0.70 & 2.07 \\
MDP & 0.26 & 0.92 & 0.24 & 0.30 & 0.36 & 2.07 \\
MSTd & 0.26 & 0.92 & 0.24 & 0.30 & 0.36 & 2.07 \\
VIP & 0.25 & 1.17 & 0.28 & 0.21 & 0.16 & 2.07 \\
LIP & 0.25 & 1.00 & 0.24 & 0.24 & 0.57 & 2.30 \\
PITv & 0.23 & 0.81 & 0.28 & 0.27 & 0.32 & 1.90 \\
PITd & 0.23 & 0.81 & 0.28 & 0.27 & 0.32 & 1.90 \\
AITv & 0.34 & 1.20 & 0.23 & 0.39 & 0.47 & 2.63 \\
MSTl & 0.26 & 0.92 & 0.24 & 0.30 & 0.36 & 2.07 \\
FST & 0.51 & 0.90 & 0.18 & 0.30 & 0.36 & 2.25 \\
CITv & 0.29 & 1.02 & 0.19 & 0.33 & 0.40 & 2.23 \\
CITd & 0.29 & 1.02 & 0.19 & 0.33 & 0.40 & 2.23 \\
7a & 0.35 & 1.24 & 0.21 & 0.41 & 0.48 & 2.68 \\
STPp & 0.29 & 1.03 & 0.18 & 0.34 & 0.40 & 2.25 \\
STPa & 0.29 & 1.03 & 0.18 & 0.34 & 0.40 & 2.25 \\
FEF & 0.22 & 0.92 & 0.35 & 0.37 & 0.35 & 2.21 \\
46 & 0.22 & 0.82 & 0.18 & 0.28 & 0.36 & 1.86 \\
TF & 0.23 & 0.66 & 0.21 & 0.24 & 0.28 & 1.62 \\
TH & 0.28 & 0.65 & 0.12 & 0.57 & 0.26 & 1.87 \\
AITd & 0.34 & 1.20 & 0.23 & 0.39 & 0.47 & 2.63 \\
\end{tabular} 
\end{center} 
\end{table} 

%% file: table_neuron_numbers.tex
\begin{table}[H] 
\begin{center} 
\begin{tabular}{llllllllll} 
\textbf{Area} & \textbf{2/3E} & \textbf{2/3I} & \textbf{4E} & \textbf{4I} & \textbf{5E} & \textbf{5I} & \textbf{6E} & \textbf{6I} & \textbf{Total} \\ 
\hline 
V1 & 47386 & 13366 & 70387 & 17597 & 20740 & 4554 & 19839 & 4063 & 197935 \\
V2 & 50521 & 14250 & 36685 & 9171 & 19079 & 4189 & 19248 & 3941 & 157087 \\
VP & 52973 & 14942 & 49292 & 12323 & 15929 & 3497 & 19130 & 3917 & 172007 \\
V3 & 58475 & 16494 & 47428 & 11857 & 12056 & 2647 & 14529 & 2975 & 166465 \\
PIP & 44343 & 12507 & 22524 & 5631 & 14742 & 3237 & 17704 & 3625 & 124318 \\
V3A & 40887 & 11532 & 23789 & 5947 & 12671 & 2782 & 15218 & 3116 & 115946 \\
MT & 60606 & 17095 & 28202 & 7050 & 14176 & 3113 & 15837 & 3243 & 149324 \\
V4t & 48175 & 13588 & 34735 & 8684 & 14857 & 3262 & 17843 & 3654 & 144801 \\
V4 & 64447 & 18178 & 33855 & 8464 & 13990 & 3072 & 14161 & 2900 & 159070 \\
PO & 44343 & 12507 & 22524 & 5631 & 14742 & 3237 & 17704 & 3625 & 124318 \\
VOT & 45313 & 12781 & 37611 & 9403 & 15828 & 3475 & 19008 & 3892 & 147315 \\
DP & 43934 & 12392 & 18896 & 4724 & 14179 & 3113 & 17028 & 3487 & 117755 \\
MIP & 41274 & 11642 & 15875 & 3969 & 7681 & 1686 & 34601 & 7086 & 123816 \\
MDP & 44343 & 12507 & 22524 & 5631 & 14742 & 3237 & 17704 & 3625 & 124318 \\
MSTd & 44343 & 12507 & 22524 & 5631 & 14742 & 3237 & 17704 & 3625 & 124318 \\
VIP & 56683 & 15988 & 26275 & 6569 & 10099 & 2217 & 7864 & 1610 & 127310 \\
LIP & 51983 & 14662 & 20095 & 5024 & 11630 & 2554 & 28115 & 5757 & 139824 \\
PITv & 45313 & 12781 & 37611 & 9403 & 15828 & 3475 & 19008 & 3892 & 147315 \\
PITd & 45313 & 12781 & 37611 & 9403 & 15828 & 3475 & 19008 & 3892 & 147315 \\
AITv & 49224 & 13884 & 18066 & 4516 & 16982 & 3729 & 20395 & 4176 & 130977 \\
MSTl & 44343 & 12507 & 22524 & 5631 & 14742 & 3237 & 17704 & 3625 & 124318 \\
FST & 36337 & 10249 & 12503 & 3126 & 12624 & 2772 & 15160 & 3104 & 95879 \\
CITv & 41696 & 11761 & 15303 & 3826 & 14385 & 3158 & 17275 & 3537 & 110944 \\
CITd & 41696 & 11761 & 15303 & 3826 & 14385 & 3158 & 17275 & 3537 & 110944 \\
7a & 49481 & 13957 & 13279 & 3320 & 15817 & 3473 & 18996 & 3890 & 122216 \\
STPp & 41677 & 11755 & 13092 & 3273 & 14218 & 3122 & 17075 & 3496 & 107712 \\
STPa & 41677 & 11755 & 13092 & 3273 & 14218 & 3122 & 17075 & 3496 & 107712 \\
FEF & 44053 & 12425 & 23143 & 5786 & 16943 & 3720 & 16128 & 3302 & 125504 \\
46 & 32581 & 9190 & 10645 & 2661 & 11850 & 2602 & 15841 & 3244 & 88617 \\
TF & 30774 & 8680 & 17143 & 4286 & 11082 & 2433 & 13310 & 2725 & 90436 \\
TH & 24712 & 6970 &  &  & 23353 & 5128 & 10861 & 2224 & 73251 \\
AITd & 49224 & 13884 & 18066 & 4516 & 16982 & 3729 & 20395 & 4176 & 130977 \\
\end{tabular} 
\end{center} 
\end{table} 

%% file: table_external_indegrees.tex
\begin{table}[H] 
\begin{center} 
\begin{tabular}{l|llllllll} 
Area & 2/3E & 2/3I & 4E & 4I & 5E & 5I & 6E & 6I  \\ 
\hline 
V1 & 1246 & 1246 & 1246 & 1246 & 1401 & 1246 & 1765 & 1246 \\
V2 & 1848 & 1848 & 1848 & 1848 & 2079 & 1848 & 2618 & 1848 \\
VP & 1756 & 1756 & 1756 & 1756 & 1976 & 1756 & 2488 & 1756 \\
V3 & 1810 & 1810 & 1810 & 1810 & 2036 & 1810 & 2564 & 1810 \\
V3A & 2703 & 2703 & 2703 & 2703 & 3041 & 2703 & 3830 & 2703 \\
MT & 2510 & 2510 & 2510 & 2510 & 2824 & 2510 & 3556 & 2510 \\
V4t & 2293 & 2293 & 2293 & 2293 & 2580 & 2293 & 3249 & 2293 \\
V4 & 2337 & 2337 & 2337 & 2337 & 2630 & 2337 & 3311 & 2337 \\
VOT & 2409 & 2409 & 2409 & 2409 & 2710 & 2409 & 3413 & 2409 \\
MSTd & 3181 & 3181 & 3181 & 3181 & 3578 & 3181 & 4506 & 3181 \\
PIP & 3327 & 3327 & 3327 & 3327 & 3743 & 3327 & 4713 & 3327 \\
PO & 3226 & 3226 & 3226 & 3226 & 3629 & 3226 & 4570 & 3226 \\
DP & 3328 & 3328 & 3328 & 3328 & 3745 & 3328 & 4716 & 3328 \\
MIP & 3474 & 3474 & 3474 & 3474 & 3908 & 3474 & 4921 & 3474 \\
MDP & 5186 & 5186 & 5186 & 5186 & 5835 & 5186 & 7348 & 5186 \\
VIP & 3378 & 3378 & 3378 & 3378 & 3800 & 3378 & 4786 & 3378 \\
LIP & 3311 & 3311 & 3311 & 3311 & 3725 & 3311 & 4691 & 3311 \\
PITv & 2441 & 2441 & 2441 & 2441 & 2746 & 2441 & 3458 & 2441 \\
PITd & 2471 & 2471 & 2471 & 2471 & 2780 & 2471 & 3501 & 2471 \\
MSTl & 3094 & 3094 & 3094 & 3094 & 3481 & 3094 & 4383 & 3094 \\
CITv & 3844 & 3844 & 3844 & 3844 & 4324 & 3844 & 5446 & 3844 \\
CITd & 3708 & 3708 & 3708 & 3708 & 4172 & 3708 & 5253 & 3708 \\
FEF & 3597 & 3597 & 3597 & 3597 & 4047 & 3597 & 5096 & 3597 \\
TF & 3805 & 3805 & 3805 & 3805 & 4280 & 3805 & 5390 & 3805 \\
AITv & 3786 & 3786 & 3786 & 3786 & 4259 & 3786 & 5364 & 3786 \\
FST & 4614 & 4614 & 4614 & 4614 & 5191 & 4614 & 6537 & 4614 \\
7a & 4361 & 4361 & 4361 & 4361 & 4906 & 4361 & 6179 & 4361 \\
STPp & 4246 & 4246 & 4246 & 4246 & 4777 & 4246 & 6015 & 4246 \\
STPa & 4032 & 4032 & 4032 & 4032 & 4536 & 4032 & 5713 & 4032 \\
46 & 4309 & 4309 & 4309 & 4309 & 4848 & 4309 & 6105 & 4309 \\
AITd & 3784 & 3784 & 3784 & 3784 & 4257 & 3784 & 5361 & 3784 \\
TH & 6590 & 5491 &  &  & 7413 & 5491 & 7780 & 5491 \\
\end{tabular} 
\end{center} 
\end{table} 